\documentclass[useAMS,usenatbib]{mn2e}

\usepackage[dvips]{graphicx}
\usepackage{natbib}
\usepackage{pifont}
\usepackage{txfonts}
\usepackage{mathrsfs}

\title[Gas phase mean opacities]{Gas phase mean opacities for varying [M/H], N/O, and C/O}
\author[Ch. Helling and W. Lucas]{Ch. Helling\thanks{E-mail: Christiane.Helling@st-and.ac.uk} and W. Lucas\\
SUPA, School of Physics and Astronomy, University of St Andrews, North Haugh, St Andrews,  KY16 9SS, UK}
\begin{document}

\maketitle
\label{firstpage}

\begin{abstract}
We present a set of gas-phase Planck mean and Rosseland mean opacity
tables applicable for simulations of star and planet formation,
stellar evolution, disk modelling at various metallicities in
hydrogen-rich environments. The tables are calculated for gas
temperatures between 1000K and 10000K and total hydrogen number
densities between $10^2$ cm$^{-3}$ and $10^{17}$ cm$^{-3}$. The
carbon-to-oxygen ratio is varied from 0.43 to well above 2.0, the
nitrogen-to-oxygen ration between 0.14 and 100.0. The tables are
calculated for a range of metallicities down to [M/H]$^{\prime}=\log
N_{\rm M}/N_{\rm H}=-7.0$.  We demonstrate how the mean opacities and
the abundances of the opacity species vary with C/O, N/O, and
[M/H]$^{\prime}$. We use the element abundances from Grevesse, Asplund
\& Sauval~(2007), and we provide additional tables for the
oxygen-abundance value from Caffau et al.(2008). All tables will be
available online under
http://star-www.st-and.ac.uk/$\sim$ch80/datasources.html.

\end{abstract}

\begin{keywords}
molecular data, radiative transfer, (stars:) planetary systems:
protoplanetary discs, stars: formation , methods: numerical
\end{keywords}

\section{Introduction}
Star and planet formation modelling (e.g. Wuchterl 2005), stellar
evolution modelling (e.g. Marques, Monteiro \& Fernandes 2008,
VandenBerg et al. 2008), galactic evolution modelling (e.g. Moll{\'a}
et al. 2007) remain dependent on the availability of opacity
data. Often, the mean opacities were only available on very sparse
temperature-density grids (Wuchterl 2003; priv. com.). Intermediate
mass asymptotic Giant Branch (AGB) stars develop from oxygen-rich
(C/O$<$1) into carbon-rich (C/O$>$1) and are a major source of element
enrichment of the insterstellar medium (ISM) (e.g. Gail \& Sedlmayr
1986, Dominik et al. 1990, Dorfi \& H{\"o}fner 1998, Wachter et
al. 2008). However, mass-loss formulae describing the matter return
from the AGB stars into the ISM do still model the gas opacity by a
constant values (Wachter et al. 2008).The modelling of single stars
does allow the treatment of a frequency dependent radiative transfer
also in dynamic models for carbon-rich (H\"ofner et al. 2003,
Gautschy-Loidl et al. 2004, Nowotny et al. 2005) and oxygen-rich
(Woitke 2006) pulsating AGB stars . Extremely metal-poor giant stars
with a metallicity of [M/H]$\lesssim -3.5$ (Cohen et al. 2008, Cayrel
et al. 2004) seem to provide strong constrains on the element yields
of first super-novae (SN) into the ISM (e.g. Cayrel et
al. 2004). Therefore, the need for gas-phase opacity tables for a
wider range of metallicities ([M/H]), carbon-to-oxygen ratios (C/O),
but also nitrogen-to-oxygen ratios (N/O) has grown with the new
challenge to model the evolution of star formation at high z, i.e. in
a comparably young universe with stars containing much less elements
than our Sun.  The actual determination of the element abundances
profs difficult (see Tsuji 2008) and is particular severe for the
solar values (Grevesse, Asplund \& Sauval~2007, Caffau et al. 2008)
which are used as reference throughout astronomy. Furthermore, we need
to be aware that stars in other galaxies can have an abundance patters
different to the Sun e.g. due to a different star formation history
(e.g. Koch et al. 2008). Helling, Winters \& Sedlmayr (2000) presented
the differences in the gas-phase mean opacities for the small and the
large Magellanic cloud which element abundances clearly differ from
that of the Sun. Based on this early work on mean opacities for AGB
star wind models, we provide an extensive grid of mean opacity tables
with an increased number of line opacity species
(Sect.~\ref{ss:input}) for varying [M/H]$^{\prime}$, C/O, and N/O
(Sect.~\ref{ss:grid} and ~\ref{s:results}). We investigate the
influence of these parameters on the Planck and the Rosseland mean
gas-phase opacities, and the influence of the oxygen abundance values
on our mean opacities stimulated by the ongoing discussion
(Sect.~\ref{s:results}).

\section{Method}

\subsection{Mean opacity formulae}
We compute mean opacity tables in the Rosseland mean approximation,
\begin{equation}
(\kappa_{\rm Ross})^{-1}=\sum_{\rm i}^{\rm N}  \Big\{ 
                       \frac{  \int_0^{\infty} 
                         \frac{1}{n_{\rm i}(\rho, T)\, \kappa^{\rm i}_{\nu}(T)}
                         \frac{\partial B_{\nu}(T)}{\partial T}\,d\nu}
                        {\int_0^{\infty} \frac{\partial B_{\nu}(T)}{\partial T} 
                        \,d\nu}
\Big\},
\label{eq:Ross}
\end{equation}
and the Planck mean approximation,
\begin{equation}
\kappa_{\rm Planck}=\sum_{\rm i}^{\rm N}  n_{\rm i} (\rho, T) \,\Big\{  
                 \frac{\int_0^{\infty} \kappa^{\rm i}_{\nu}(T)\,B_{\nu}(T)\,d\nu}
                      {\int_0^{\infty} \,B_{\nu}(T)\,d\nu}\Big\},
\label{eq:Planck}
\end{equation}
with $B_{\nu}(T)$ the Planck function at frequency $\nu$ and local
temperature $T$ in [K]. $n_{\rm i}$ is the number density in
[cm$^{-3}$] of the gas-phase opacity species, $\rho$ is the local gas
density in [g\,cm$^{-3}$]. The frequency-dependent absorption
coefficient $\kappa^{\rm i}_{\nu}(T)$ is derived from pre-tabulated,
opacity-sampled line lists for each of the gas-phase opacity source i
($\rm i=1\ldots N$). The integrals in Eqs.~\ref{eq:Ross}
and~\ref{eq:Planck} are evaluated at 5608 sampling points in frequency
space which are distributed across the entire wavelength interval
$0.2\mu$m$\,\ldots\,20\mu$m. This distribution samples densest the
wavelength ranges were the local Planck functions between
1000K$\,\ldots\,$6000K peak. These Planck functions are meant to cover
a range of local temperatures where molecules are important opacity
sources. For more details see Helling \& J{\o}rgensen (1998). Helling
\& J{\o}rgensen (1998) demonstrated that the temperature-pressure
structure of M-stars is not affected by the number of sampling points
for which the radiative transfer is evaluated if changed from
$\sim$22\,000, to 5608, to $\sim$500 points. Differences in the
temperature-pressure structures become apparent for carbon-rich models
and for low-metallicity but only if the sampling number is as low as
$\sim$500. Carbon-rich model atmospheres and low-metallicity model
atmospheres show generally a higher standard deviation with decreasing
number of sampling points well below 5000 if looked at the changes in
the temperature-pressure structure. However, the evaluation of
the integrals in Eqs.~\ref{eq:Ross} and~\ref{eq:Planck} will to some
extend depends on the integral dicretization, and hence, on the number of
frequency points taken into account.

 \subsection{Gas opacity and gas-phase chemistry data}\label{ss:input}

The number density of the gas-phase opacity species, $n_{\rm i}$, is
computed for a given temperature, T [K], and total hydrogen number
density, n$_{\rm H}$ [cm$^{-3}$], assuming chemical equilibrium for 14
elements (H, He, C, N, O, Si, Mg, Al, Fe, S, Na, K, Ti, Ca) and 158
molecules with equilibrium constants fitted to the thermodynamical
molecular data of the electronic version of the JANAF tables (Chase et
al. 1986). The equilibrium constant for TiC are from Gauger et
al. (see Helling, Winters \& Sedlmayr 2000), for CaH from Tsuji (1973), and FeH from
Burrows (2008, priv. com.). First ionisation states of the elements
are calculated. The element abundance data are those of Grevesse,
Asplund \& Sauval (2007), 
%\citet*{gas}, 
and are summarised in Table~\ref{abundt}, otherwise will be stated (see
Sect.~\ref{ss:oxygen}). The element abundance are adjusted according to
the metallicity [M/H]$^{\prime}$, the C/O, or the N/O if needed
(Sect.~\ref{ss:grid}).

We include continuum opacity sources (HI (Karzas \& Latter 1961), H-
(John 1988), H+H (Doyle 1968), H$_2^-$ (Somerville 1964), H$_2^+$
(Mihalas 1965), HeI (Peach 1970), He- (Carbon et al. 1969), and CI,
MgI, AlI, and SiI (Peach 1970). Thomson scattering and Rayleigh
scattering for HI and HeI (Dalgarno 1962)). The line opacity sources
are taken from the Copenhagen SCAN data base (CO (Goorvitch 1994), TiO
(J{\o}rgensen 1994b), SiO (Langhoff \& Bauschilder 1993), H$_2$O
(J{\o}rgensen \& Jensen 1993), CH (J{\o}rgensen et al. 1996), CN
(J{\o}rgensen \& Larsson 1990), C$_2$ (Querci, Querci \& Tsuji 1974), C$_3$
(J{\o}rgensen et al. 1989), HCN and C$_2$H$_2$ (J{\o}rgensen 1990 in
an updated version), H$_2$-H$_2$ and H$_2$-He collision induced
absorption from Borysow (2002 and priv. com.), and were extended
(Schnabel 2001) by the HITRAN data base (CH$_4$, NH$_3$, CO$_2$,
SO$_2$, NO$_2$, NO, OH, N$_2$; Rothman et al. 2003). Line list
from all these data were opacity sampled according to Helling \&
J{\o}rgensen (1998).  

In the most general sense, the opacity data can only be as good as the
line-list data are. To access the quality of line lists, studies were
undertaken to which extant different line lists can help to fit
observed spectra.  For example, Jones et al. (2002) used {\sc Phoenix}
model atmosphere structures and different water line lists to produce
synthetic M-dwarf spectra in comparison to observations.  
%They
%conclude that the SCAN H$_2$O line list overestimates the water
%absorption in M-dwarf spectra compared to the NASA Ames line list by
%Patridge \& Schwenke (1997).  
J{\o}rgensen (2003) discusses, amongst other molecules, the
completeness of H$_2$O-line list data, and its implications for
comparisons as in Jones et al. (2002).  In addition to the line lists
used in this paper, various other line lists are available also for
TiO (Plez 1998, Schwenke 1998) and HCN (Harris et al. 2002, 2008; see
also J{\o}rgensen 2005). The HITRAN spectroscopic data are regularly
updated (Rothman et al. 2005) but the data used cover a maximum
temperature interval of $\sim 400\,\ldots\sim 3000$K only. They were
initially compiled for planetary systems. Much work is done to compile
more complete line lists, one example being CH$_4$ (Borysow et
al. 2003, Nassar \& Bernath 2003, Rothman et al. 2005) for which
Homeier, Hauschildt \& Allard (2003) present a line list data
comparison. It therefore needs to be appreciated that the line-list
date we are using are incomplete. We tested the effect of missing weak
lines and uncertainties thereof (e.g line strength) on the calculation
of mean opacities. We artificially increased the line opacity for
lines below a certain threshold. Two cases are shown in
Fig.~\ref{old_and_linelist}: (i) all wavelength with
$\kappa(\lambda)<10^{-2}$cm$^2$/g were enhanced by a factor of 3, and
(ii) all wavelength with $\kappa(\lambda)<10^{-5}$cm$^2$/g were
enhanced by a factor of 10.  These thresholds are small compared to
the strongest H$_2$O or TiO line of $\kappa(\lambda)=1 - 10$cm$^2$/g
at e.g. T$\approx 2000$K and $\rho\approx
10^{-9}$g/cm$^3$. Figure~\ref{old_and_linelist} demonstrates that
possibly incomplete line lists have the largest effect on the
Rosseland mean opacities since it is dominated by transparent spectral
regions due to the harmonic nature of the averaging process
(Eq.~\ref{eq:Ross}). Therefore, uncertainties in weak-line opacities
do have a considerable influence in particular on the Rosseland mean
opacities for a tentative line opacity threshold of
$\kappa(\lambda)\lesssim 10^{-2}$cm$^2$/g. At temperatures below 1000K
uncertainties in the line opacities become noticiable already at a
level of $\kappa(\lambda)<10^{-5}$cm$^2$/g.

 \begin{figure*}
\begin{tabular}{cc}
\hspace*{-0.6cm}oxygen-rich \hspace*{2.0cm}carbon-rich &\hspace*{-0.6cm} oxygen-rich \hspace*{2cm} carbon-rich\\[-2.2cm]
  \vspace{-2cm}\hspace{-1.0cm}\includegraphics[width=100mm]{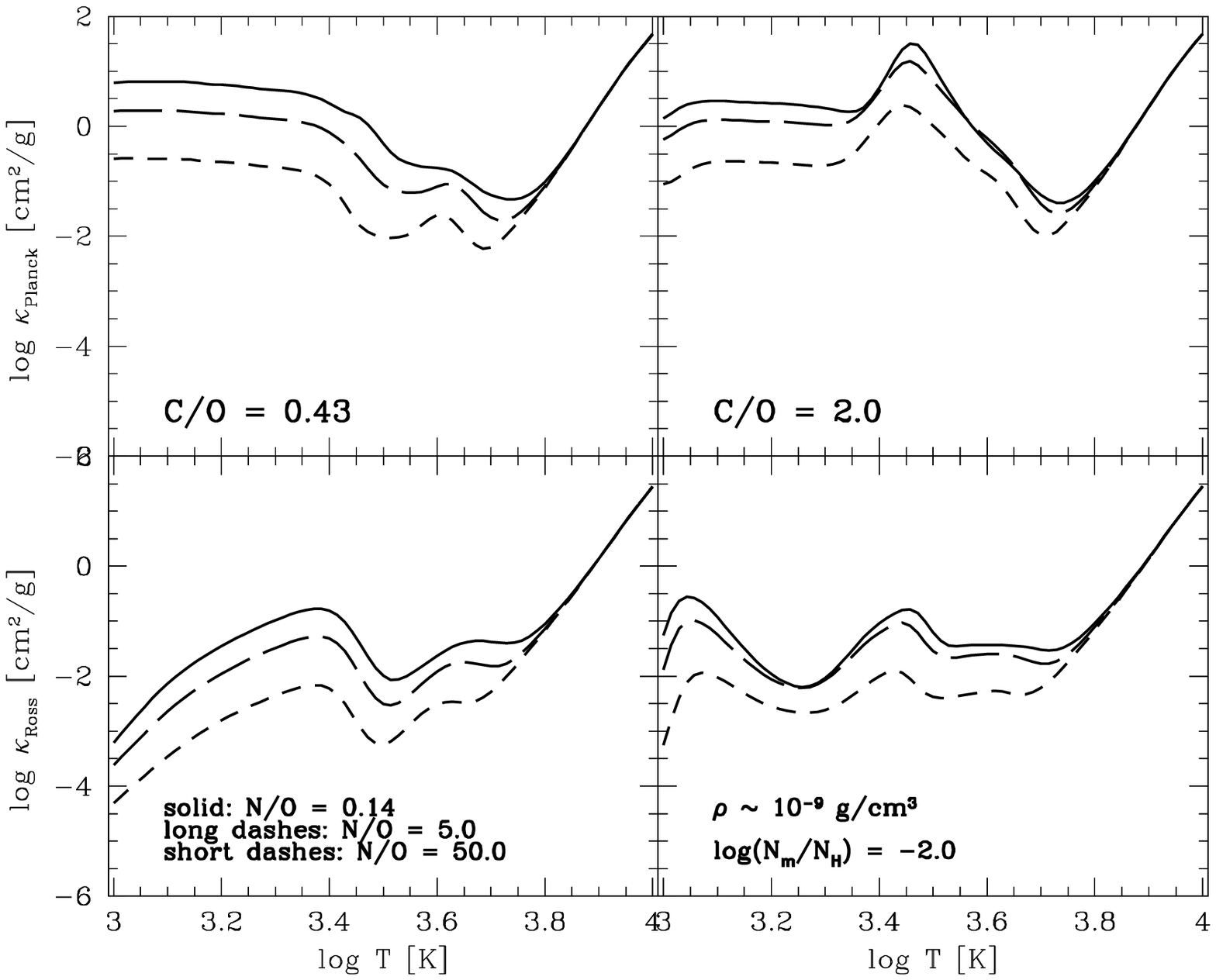}&\hspace{-1.2cm}\includegraphics[width=100mm]{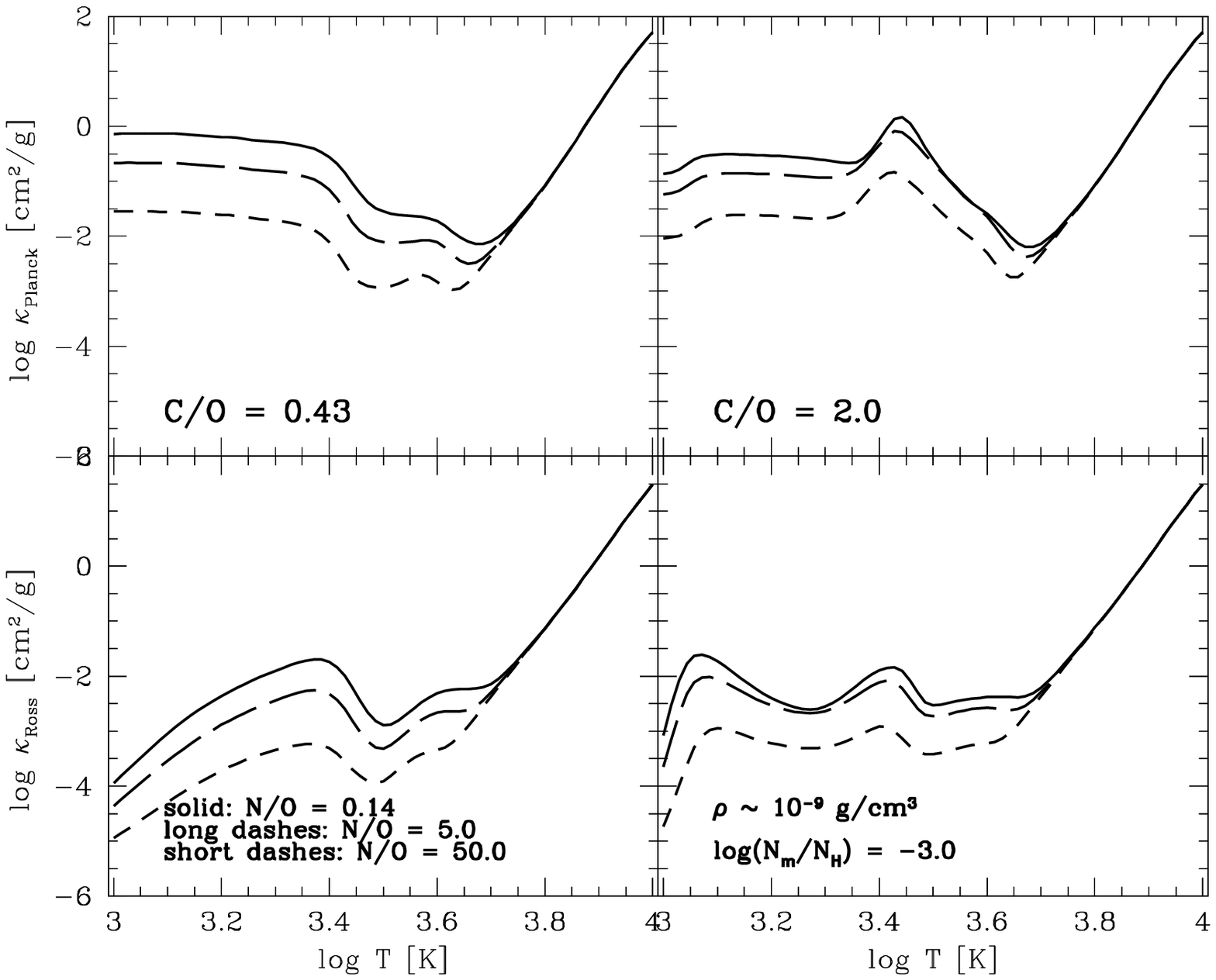}\\[-1.2cm]
  \vspace{-2.7cm}\hspace{-1.0cm}\includegraphics[width=100mm]{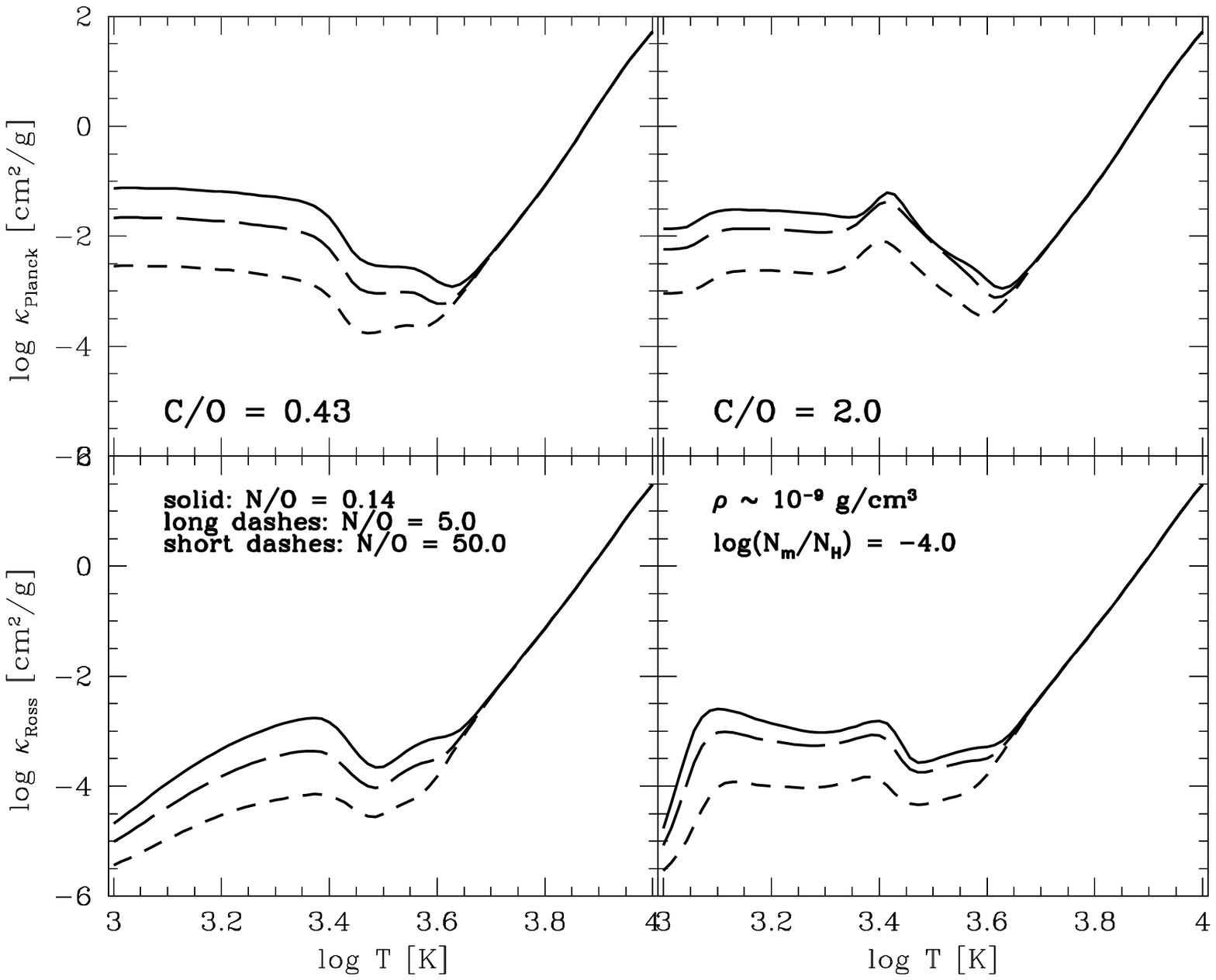}&\hspace{-1.2cm}\includegraphics[width=100mm]{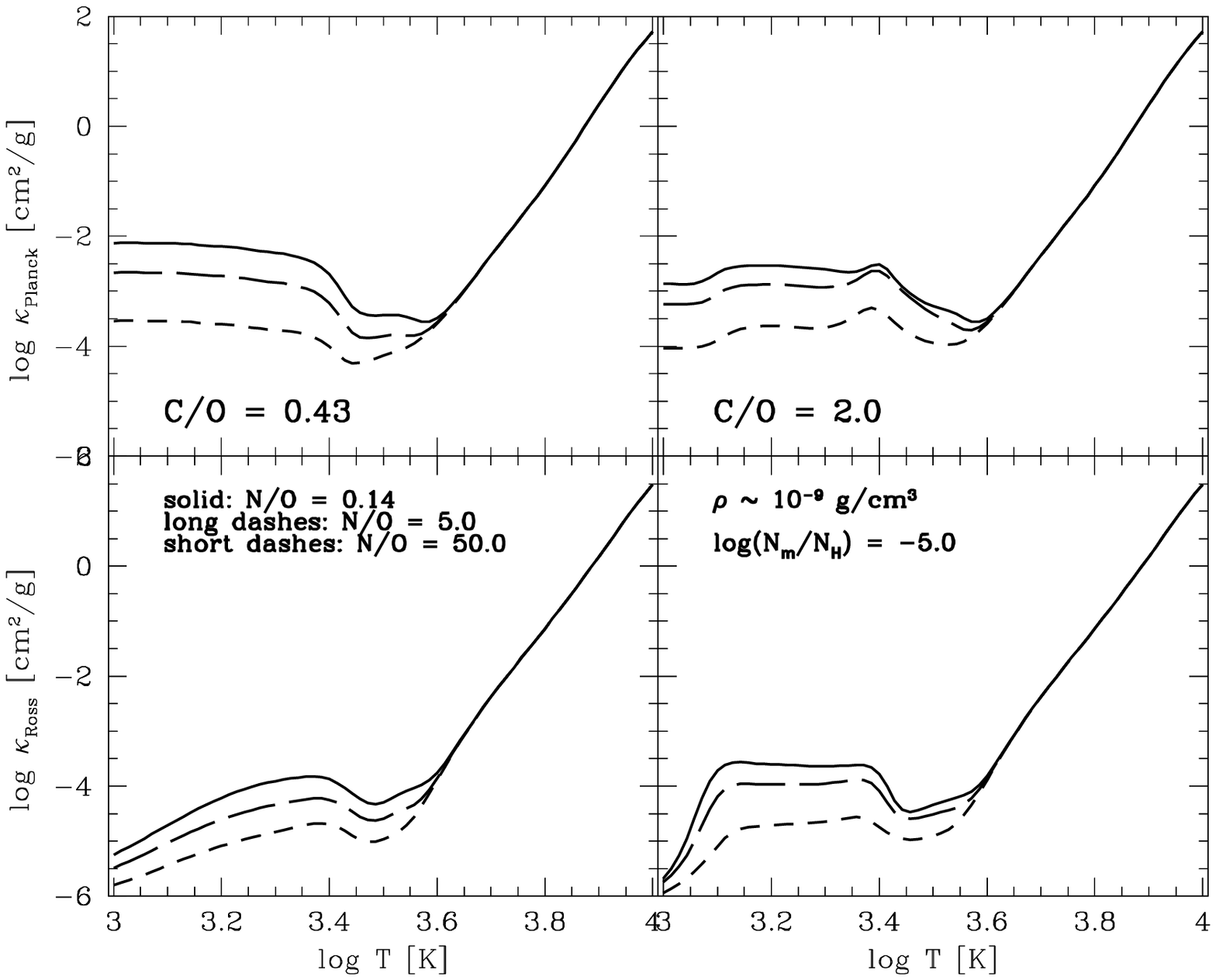}\\[-0.5cm]
  \vspace{-2.7cm}\hspace{-1.0cm}\includegraphics[width=100mm]{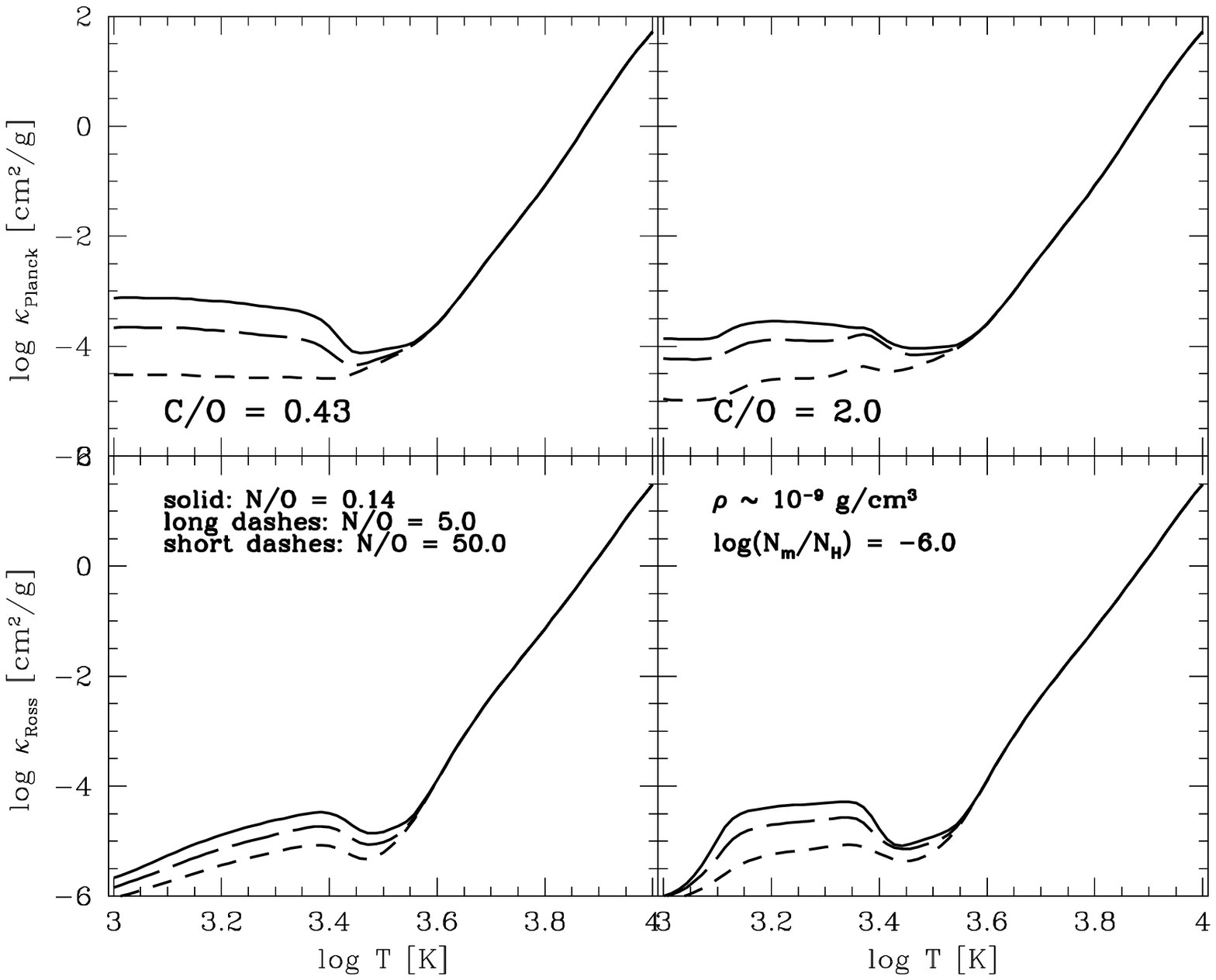}&\hspace{-1.2cm}\includegraphics[width=100mm]{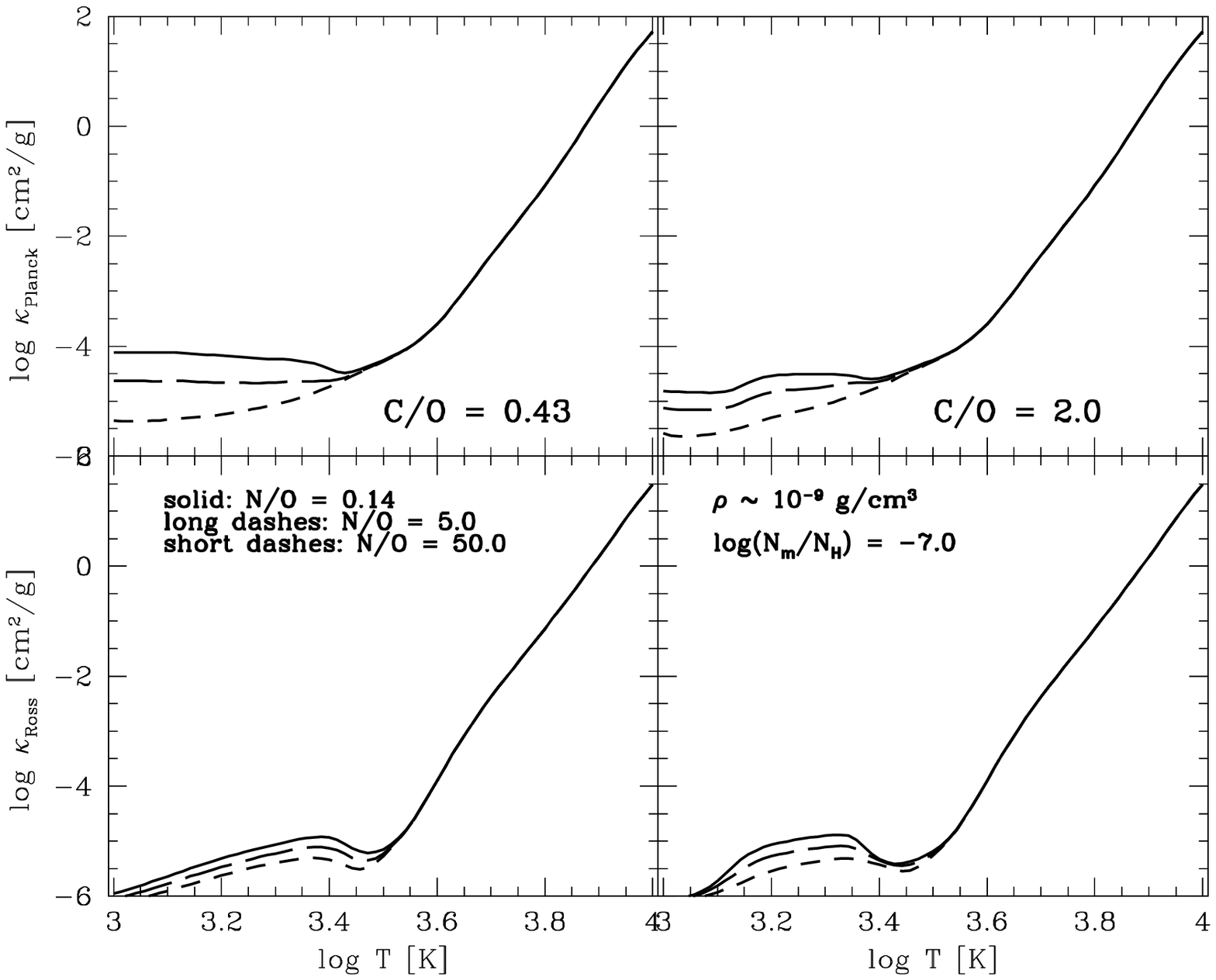}\\[2cm]
\end{tabular}
  \caption{Oxygen-rich (C/O=0.43) and carbon-rich (C/O=2.0) gas-phase mean
   opacities for varying nitrogen abundance (N/O=0.14, 5.0,
   50.0). Each set of four plots corresponds to a different 
   metallicity, with values $[\rmn{M/H}]^{\prime} = $ -2.0, -3.0, -4.0, -5.0, -6.0 and -7.0.
   The data is shown for a hydrogen number density $n_\rmn{H} = 1.931 \times 10^{15} \rmn{cm^{-3}}$
   ($\rho \sim 10^{-9} \rmn{g cm^{-3}}$).}
   
%   parameters were changed, so was the overall density corresponding to $n_\rmn{H}$,
%   but the value was roughly $\rho \sim 10^{-9}$ g $\rmn{cm^{-3}}$. As is expected,
%   the Rosseland mean is generally smaller than the Planck mean.}
  
  \label{tables} 
 \end{figure*}

\begin{figure*}
\begin{tabular}{cc}
high metallicity & low metallicity\\[-2.2cm]
  \hspace{-1.3cm}\includegraphics[width=100mm]{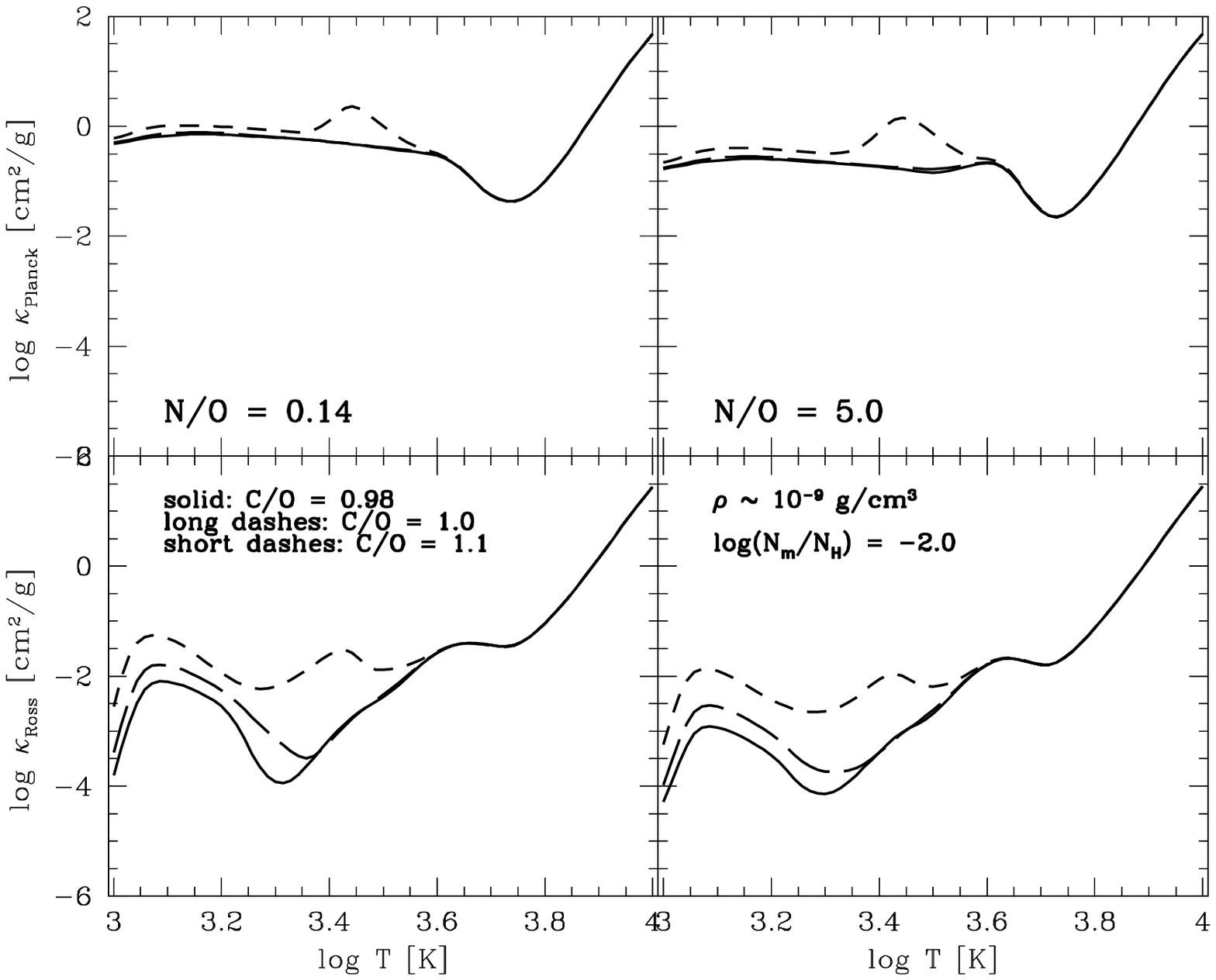} &\hspace{-1cm}\includegraphics[width=100mm]{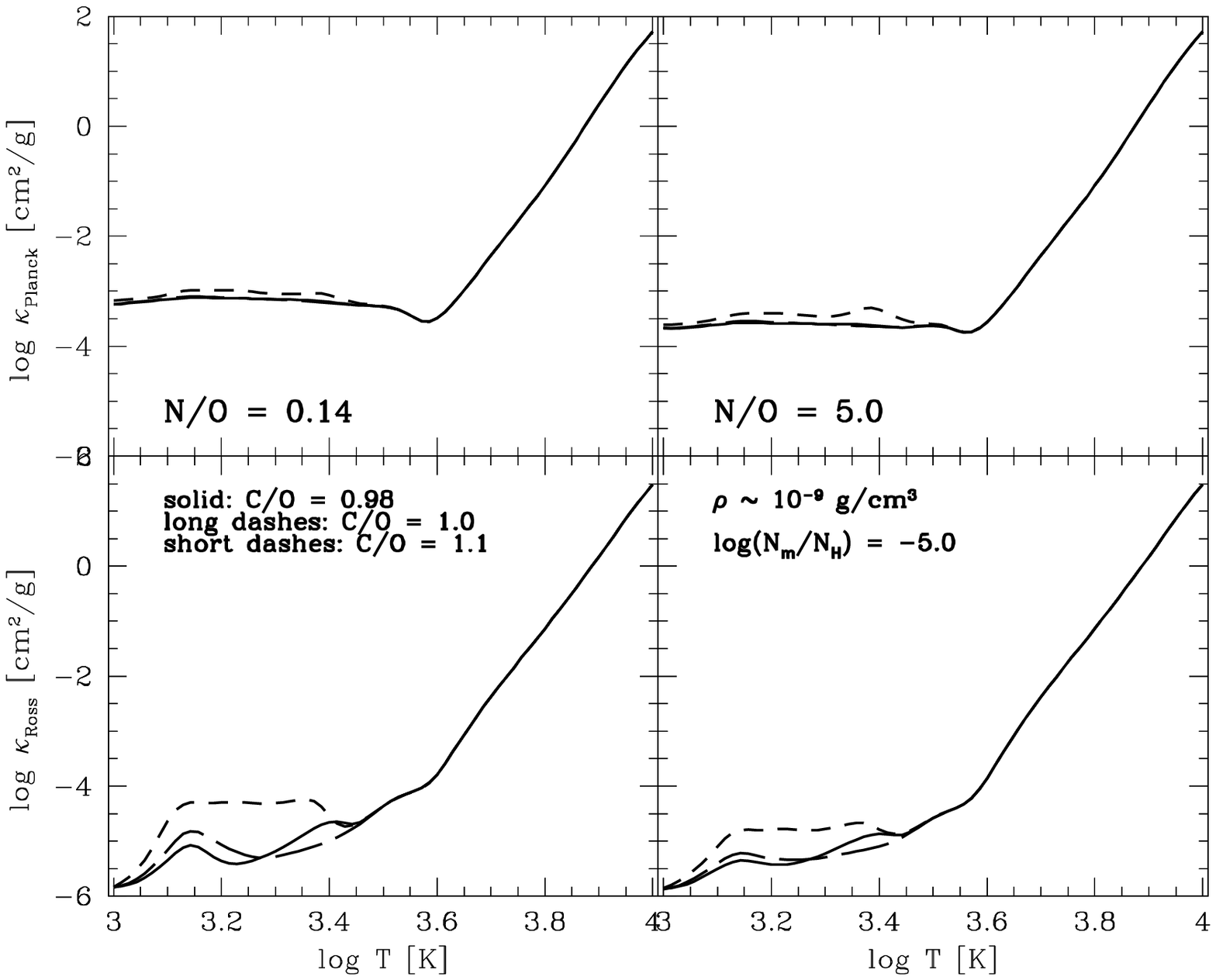}\\[-0.8cm]
\end{tabular}
  \caption{Gas-phase mean opacities for the carbon-to-oxygen-rich transition at
   C/O$\,\approx\,1$. The left set shows a high metallicity case
   ($[\rmn{M/H}]^{\prime} = $ -2.0), and the right a low metallicity case
   $[\rmn{M/H}]^{\prime} = $ -5.0) both for N/O=0.14, 5.0 for $n_\rmn{H} =
   1.931 \times 10^{15}$ $\rmn{cm^{-3}}$.}
  \label{counityzoom}
 \end{figure*}

 \subsection{Grid values}\label{ss:grid}
All mean opacities are calculated for a grid of $71\times 71$
  temperatures, T, and hydrogen number densities, $n_\rmn{H}$. The
  values of $T$ running from 1000K to 10000K, and $n_\rmn{H}$ running
  from $1.0\,10^{17} \rmn{cm}^{-3}$ to $1.0\, 10^2 \rmn{cm}^{-3}$. We
  calculate mean opacities for varying input element abundances. We
  vary the C/O ratio, a value which drastically alters the system's
  chemistry as it passes unity, the N/O ratio which also affects the
  abundance of the resulting molecular species, and the metallicity
  [M/H]$^{\prime}$.  Carbon in particular is important here, as it may
  well be from low/intermediate mass stars which pass through the AGB
  phase that the majority of carbon is synthesised (Chiappini, Romano
  \& Matteucci 2003).  %\citep*{ocnevo}.  The nitrogen abundances are
  of particular interest for modelling high-mass stars (S. Campbell,
  priv. com.).
 
    \begin{table}
  %\begin{minipage}{27mm}
   \caption{Element abundances, $\epsilon_\rmn{el}=\epsilon^{\rm GrAsSa}_\rmn{el}=$, from Grevesse,
    Asplund \& Sauval~(2007) for calculating the gas-phase chemistry.}
   \label{abundt}
%   \begin{tabular}{lr@{.}lr@(.)l}
   \begin{tabular}{lclc}
    \hline
    Element & $\epsilon_\rmn{el}$ & Element & $\epsilon_\rmn{el}$ \\
    \hline
    H  & 12.00  & Al & 6.37\\
    He & 10.93  & Si & 7.51\\
    C  & 8.39   & S  & 7.14\\
    N  & 7.78   & K  & 5.08\\
    O  & 8.66   & Ca & 6.31\\
    Na & 6.17   & Ti & 4.90\\ 
    Mg & 7.53   & Fe & 7.45\\
    \hline
   \end{tabular}
  %\end{minipage}
  \end{table}
  
  The abundance of an element is taken to be $\epsilon_\rmn{el} =
  \rmn{log}(N_\rmn{el}/N_\rmn{H}) + 12.0$ (ie. $\epsilon_\rmn{H} =
  12.0$). Then, the ratio of the element in question (here: C, N or
  all if metallicity is changed) to oxygen can in turn be given as
  $\rmn{el/O} = 10^{\epsilon_\rmn{el}-\epsilon_\rmn{O}}$.  We change
  the table values (Table \ref{abundt}) for $\epsilon_\rmn{C}$ and
  $\epsilon_\rmn{N}$ to give the desired C/O and N/O ratios, for
  example setting $\epsilon_\rmn{C} = \epsilon_\rmn{O} + \rmn{log}
  (\rmn{C/O})$. This also changes the overall metallicity. We define
  the metallicity as $[\rmn{M/H}]^{\prime} =
  \rmn{log}(N_\rmn{M}/N_\rmn{H})$, where $N_\rmn{M}$ is the total
  number density of all the atoms heavier than He. Therefore,
  $[\rmn{M/H}]^{\prime} = 0.0$ is the metallicity such that the total
  number of metal atoms equals the number of hydrogen atoms.  Hence,
  for a desired metallicity each metal must be modified by
  $\epsilon_\rmn{el}^\rmn{new} = \epsilon_\rmn{el}^\rmn{old} +
  \rmn{log}(\beta)$ with $\beta = 10^{\rmn{[M/H]^{\prime}
  }}\cdot(N_\rmn{M}^\rmn{old}/N_\rmn{H})^{-1}$. The same applies if we
  vary the C- and the N-abundance (M=C, N) values only.
Note that our metallicity is {\it not} given relative to the solar
  values but is an absolute value. The relative metallicity values,
  [M/H], would write [M/H]=[M/H]$^{\prime} -$[M/H]$_{\odot}$ with
  [M/H]$_{\odot}$ the solar metallicity (see Sect.~\ref{ss:oxygen}).

  The grid of values for the metallicity and C/O and N/O ratios are
  shown in Table \ref{paragrid}.
  \begin{table}
   \caption{Grid values of C/O, N/O and [M/H]$^{\prime}$ for mean opacity tables.}
   \label{paragrid}
   \begin{tabular}{@{}lc}
    \hline
    C/O & 0.43, 0.80, 0.90, 0.98, 1.0, 1.1, 2.0, 10.0, 50.0, 100.0 \\
    N/O & 0.14, 0.50, 1.0, 2.0, 5.0, 50.0, 100.0 \\
    $[\rmn{M/H}]^{\prime}$ & $-2.0$, $-3.0$, $-4.0$, $-5.0$, $-6.0$, $-7.0$ \\
    \hline
   \end{tabular}
   \medskip
  \end{table}
  The large number of C/O and N/O values should allow models to use
  the opacity tables for a number of different purposes, like star and
  planet formation or disk simulations. 
%The close intervals of C/O
%  around 1 should also shed some light on how the chemistry of the
%  system changes as carbon becomes more abundant than oxygen.

 We note, that our opacity tables are calculated for a
hydrogen-rich environment (with the solar abundances as reference) in
contrast to the helium-rich environment of e.g. white dwarfs. Hence,
we have not changed the relative abundance of hydrogen to helium, and
we also refrain from adopting a non-ideal equation of state. For example,
Harris et al. (2004) demonstrate the influence of HeH$^+$ on the
opacity of a helium-rich atmosphere. HeH$^+$ is, hence, suggested as
an important opacity source for metal-free Population III stars. Also
H$_3^+$ was appreciated as candidate for efficient cooling in
metal-free gases which is, however, not included in our mean opacity
calculation.

\section{Results}\label{s:results}

\begin{figure}
%   \hspace{-0.5cm}
   %\vspace{-2cm} 
\includegraphics[width=85mm]{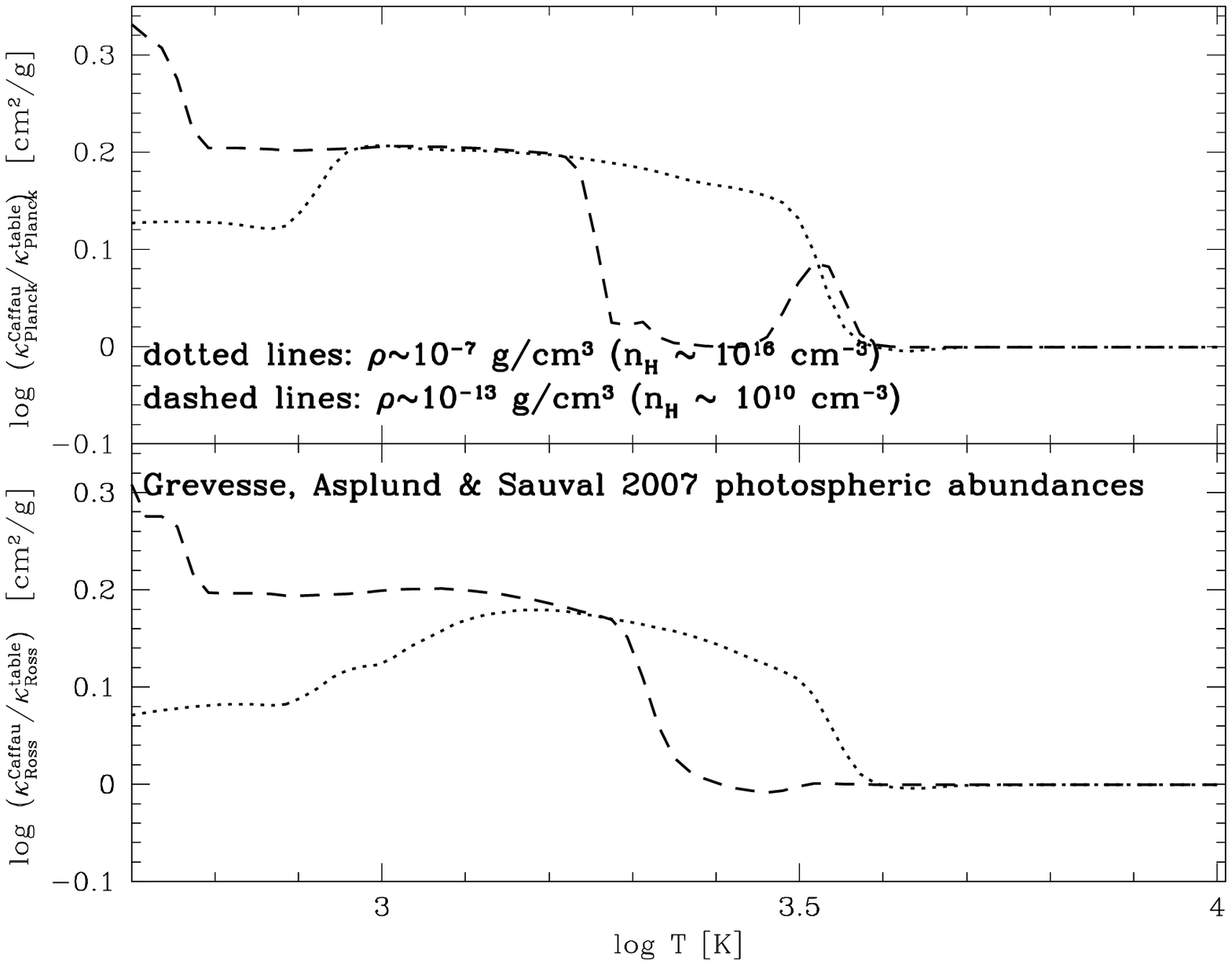}\\
\includegraphics[width=85mm]{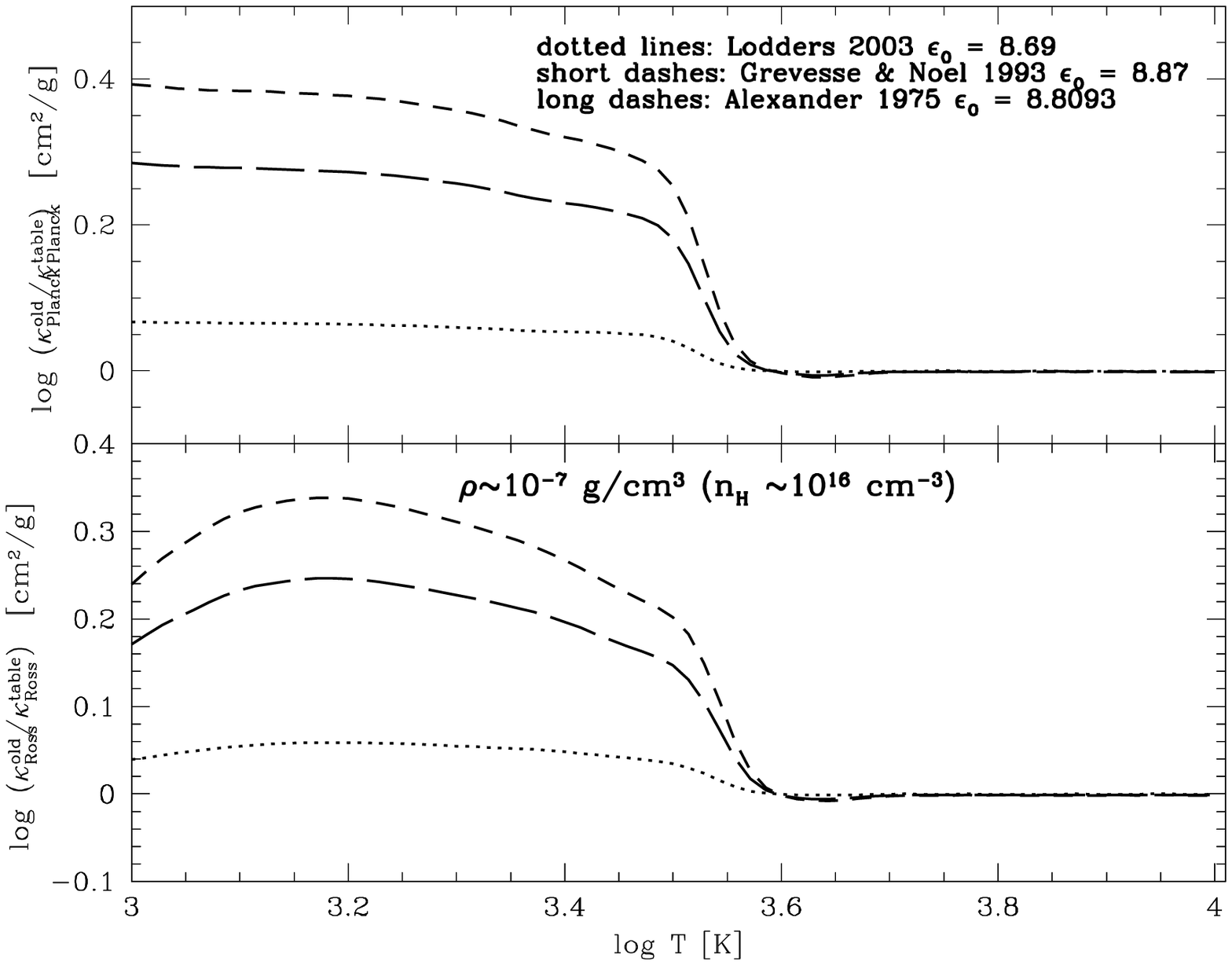}
%   \vspace{-1cm}
   \caption{Differences in Planck mean and Rosseland mean (top panel)
    opacities using different oxygen abundance (Grevesse, Asplund \&
    Sauval~2007 $\epsilon^{\rm GrAsSa}_\rmn{O} = 8.66$; Caffau et al. 2008
    $\epsilon^{\rm Caffau}_\rmn{O} = 8.76$). The logarithmic ratio of the opacities
    is plotted against temperature. In the lower panel, differences in the mean
    opacities are also shown for the oxygen abundance recommended by Lodders
    (2003, $\epsilon_\rmn{O} = 8.69$), Grevesse \& Noel (1993,
    $\epsilon_\rmn{O} = 8.87$, used in Helling, Winters \& Sedlmayr 2000), and
    Alexander (1975, $\epsilon_\rmn{O} = 8.8093$). Note that changing
    the oxygen abundance results in changes of C/O, N/O and
    [M/H]$^{\prime}$. See text for details.}
   \label{caffgasratio}
   \medskip
  \end{figure}
  
\begin{figure}
\includegraphics[width=85mm]{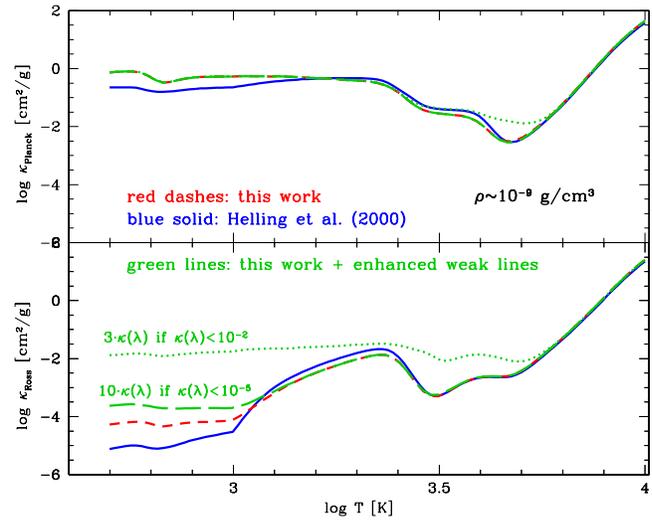}\\
%   \vspace{-1cm}
   \caption{Comparing mean opacities of this work (red dashes) to
   Helling, Winters \& Sedlmayr (2000, blue solid). In comparison, the
   effect of weak-line opacities is shown for two cases (green dots
   and green long dashes; for details see Sect.~\ref{ss:input}).} 
   \label{old_and_linelist}
   \medskip
  \end{figure}

 \subsection{Opacity tables}
 Figure \ref{tables} shows plots of several opacity tables for
 different combinations of the C/O and N/O ratios, and the
 metallicity (Table \ref{paragrid}). The effects of increasing the N/O
 ratio and decreasing the metallicity are clearly seen to produce an
 overall reduction in the mean opacities. The effect due to
 metallicity can be explained as the fewer metals are present, the
 fewer complex molecules can form, hence, decreasing absorption coefficient.
% Several noticeable features in the plots are  reduced by
% decreasing the metallicity.
The effect of changing metallicities on the mean opacity values
dominates over the changes in C/O or N/O.
 
 One interesting change is seen in the carbon-rich Rosseland mean. The
 two strong opacity peaks in the lower temperature region grow smaller
 as the metallicity is lowered. Eventually, however, they, and the
 trough between them, form a distinctive plateau by $[\rmn{M/H}]^{\prime} =
 -5.0$. In the ultra-low metallicity cases of $[\rmn{M/H}]^{\prime} = -6.0,
 -7.0$, the opacities' features are lessened even more, so that by
 $[\rmn{M/H}]^{\prime}=-7.0$ both the Planck and Rosseland means are very low
 and almost constant before the increase due to the continuum opacity
 sources begins to dominate at about $\rmn{log}(T_\rmn{g}) \sim
 3.5$. This increase at high temperature is seen in all data, but the
 lower the metallicity the lower the temperature at which it can be
 seen due to the drop in molecular contribution to the opacity.
 
 Interesting to note is that increasing the N/O ratio also appears to
 lower both the Planck and Rosseland mean opacities considerably. It
 should however be taken into consideration that by increasing the
 nitrogen content, the other metals' abundances will be decreased
 accordingly in order to agree with the given metallicity. The changes
 by increasing the relative nitrogen abundance is complex because it
 extensively affects the chemical composition of the gas phase
 (Sect.~\ref{ss:chem}).

\subsubsection{The oxygen-to-carbon-rich transition at C/O$\approx 1$}
 The dichotomy between oxygen-rich ($\rmn{C/O} < 1$) and carbon-rich
 ($\rmn{C/O} > 1$) opacities is very noticeable. This is mainly due to
 the locking of the majority of either carbon or oxygen, whichever is
 the least abundant, in CO. This leaves comparatively little to form
 other molecules, while the more abundant of the two elements may do
 so freely. The changes that occur as C/O passes 1.0 can be seen in
 more detail in Figure \ref{counityzoom}. Here little difference is
 seen between the C/O ratios of 0.98 (solid lines) and 1.0 (long
 dashed lines), particularly in the Planck means. With these ratios,
 little of either element will be available. Meanwhile, at $\rmn{C/O}
 = 1.1$ (short dashed lines) a large difference is seen as it becomes
 possible for carbon to form opacity-source molecules. Changes in the
 nitrogen abundances do not affect the mean opacities at C/O$\approx
 1$.

 Figure \ref{counityzoom} demonstrates that the Planck mean values are
 only mildly affected by the drastic chemistry changes in the
 oxygen-to-carbon-rich transition, but the Rosseland mean values do
 change in particular at low temperatures. This is an indication that
 line opacity sources with a large number of weak lines (like OH, CH,
 CN or HCN) are important opacity carrier here. The dramatic effect on
 the atmosphere structure for the oxygen-to-carbon-rich transition has
 been discussed for instance in J{\o}rgensen (1994a).

 \begin{figure*}
  \hspace{-1cm}\includegraphics[width=64mm]{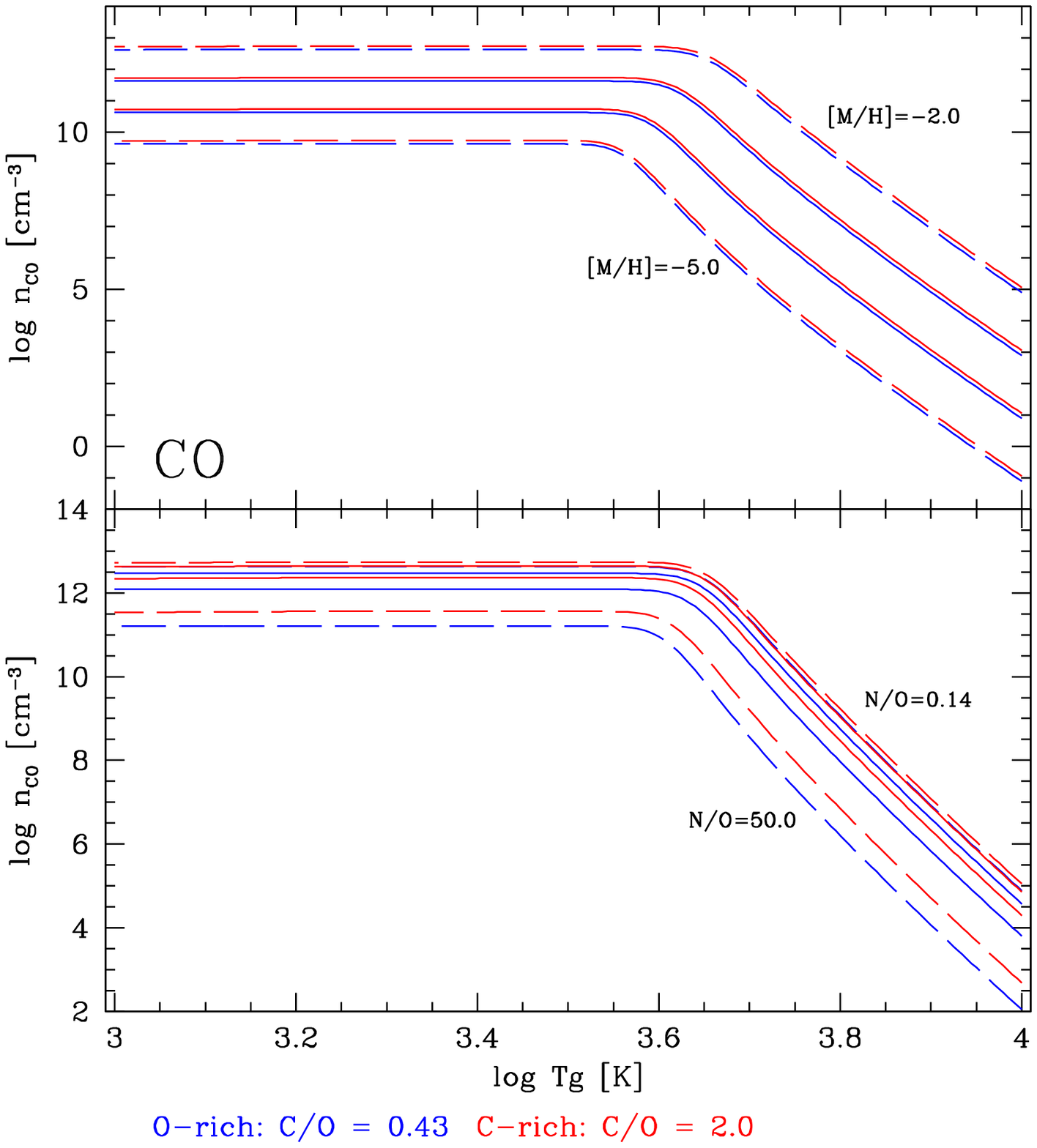}\hspace{-1.2cm}\includegraphics[width=64mm]{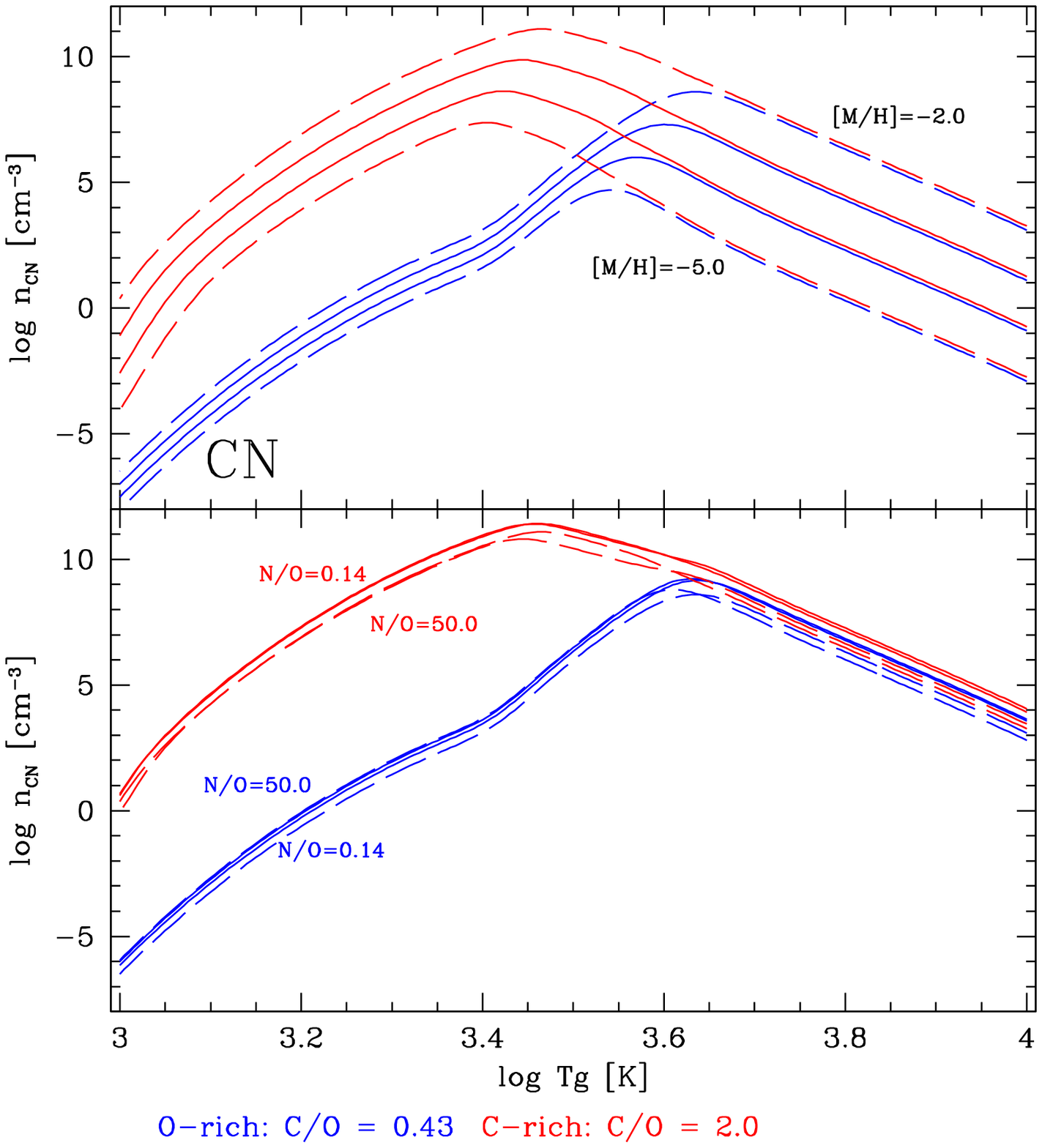}\hspace{-1.2cm}\includegraphics[width=64mm]{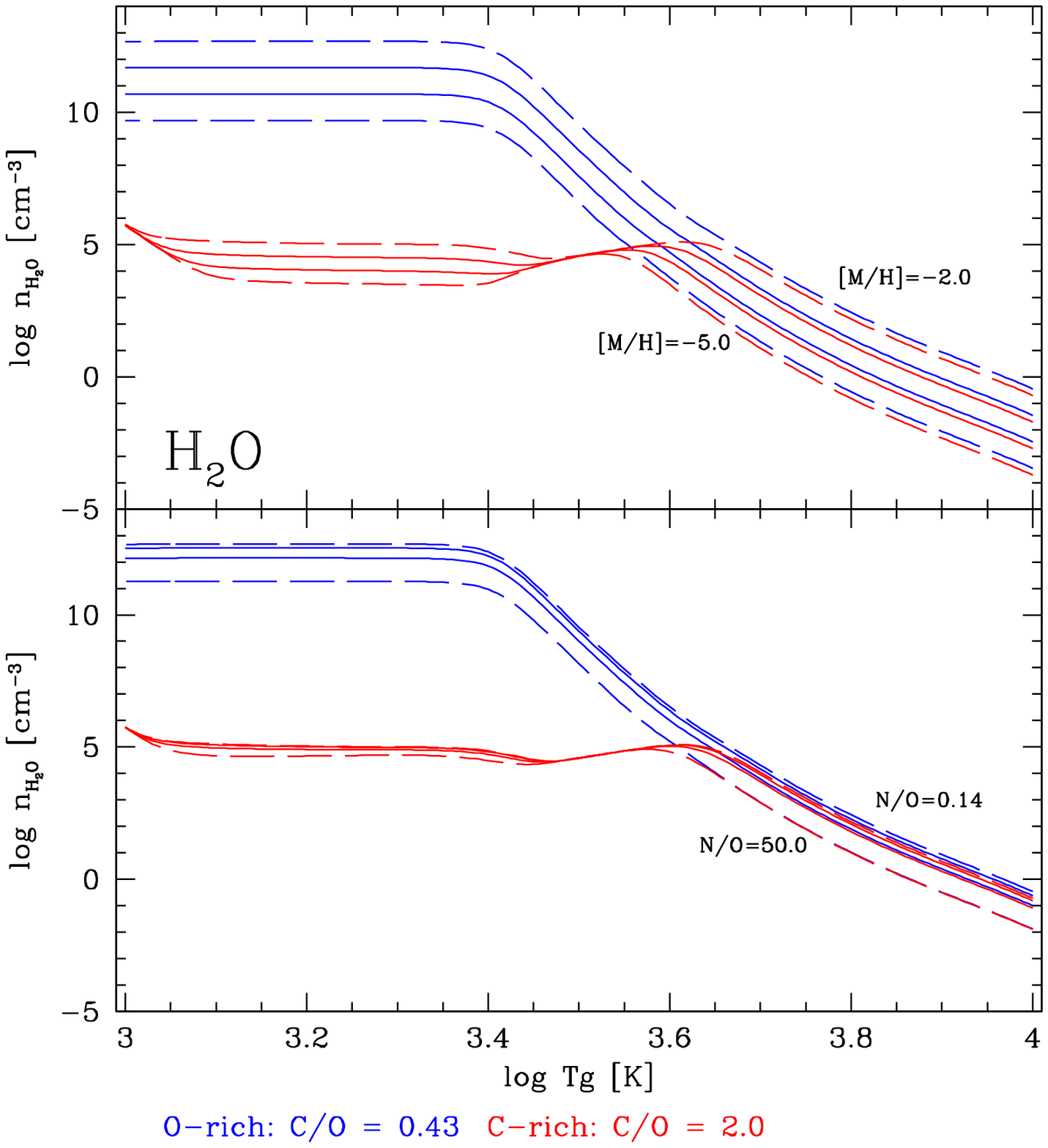}\\
  \hspace{-1cm}\includegraphics[width=64mm]{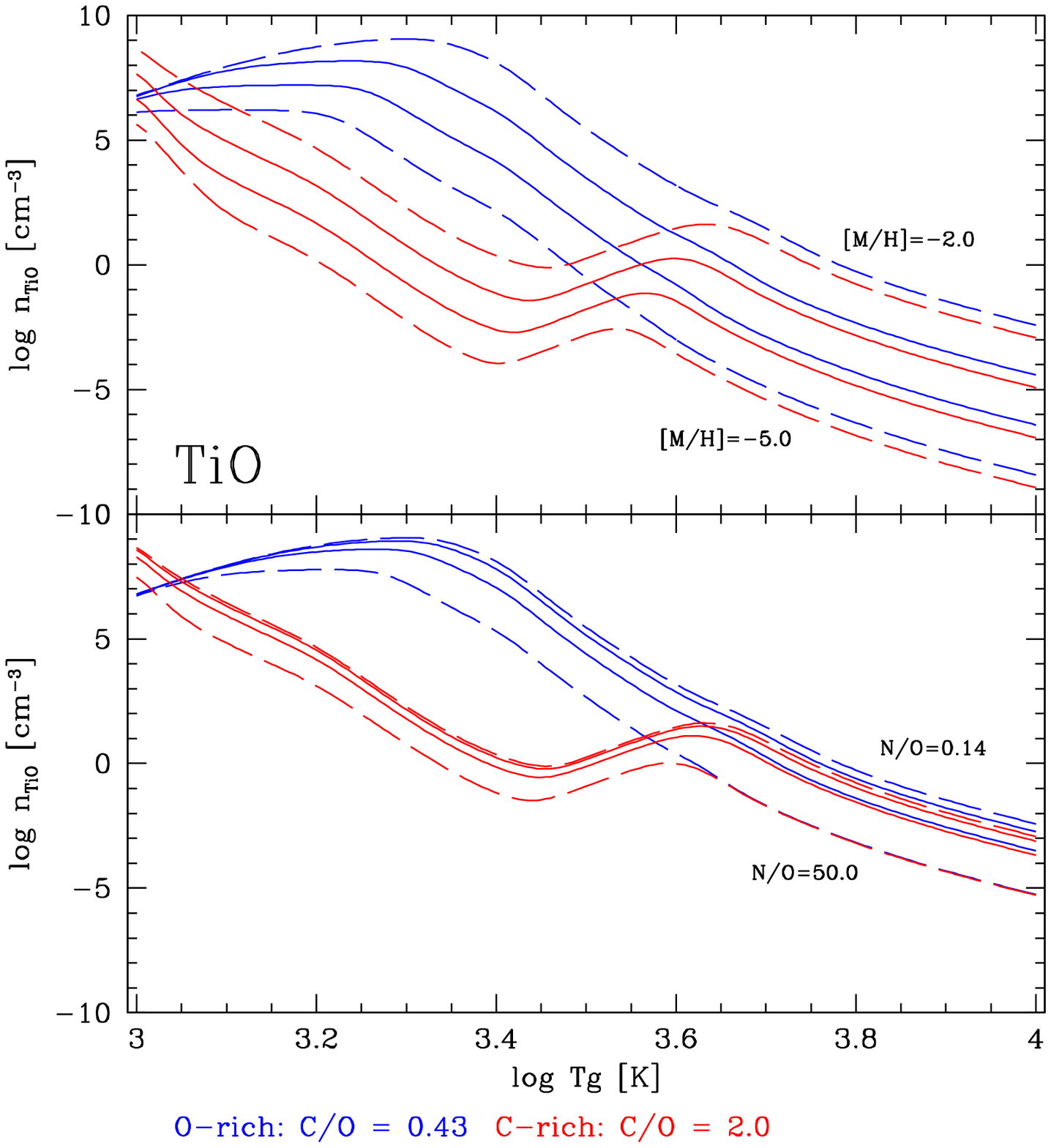}\hspace{-1.2cm}\includegraphics[width=64mm]{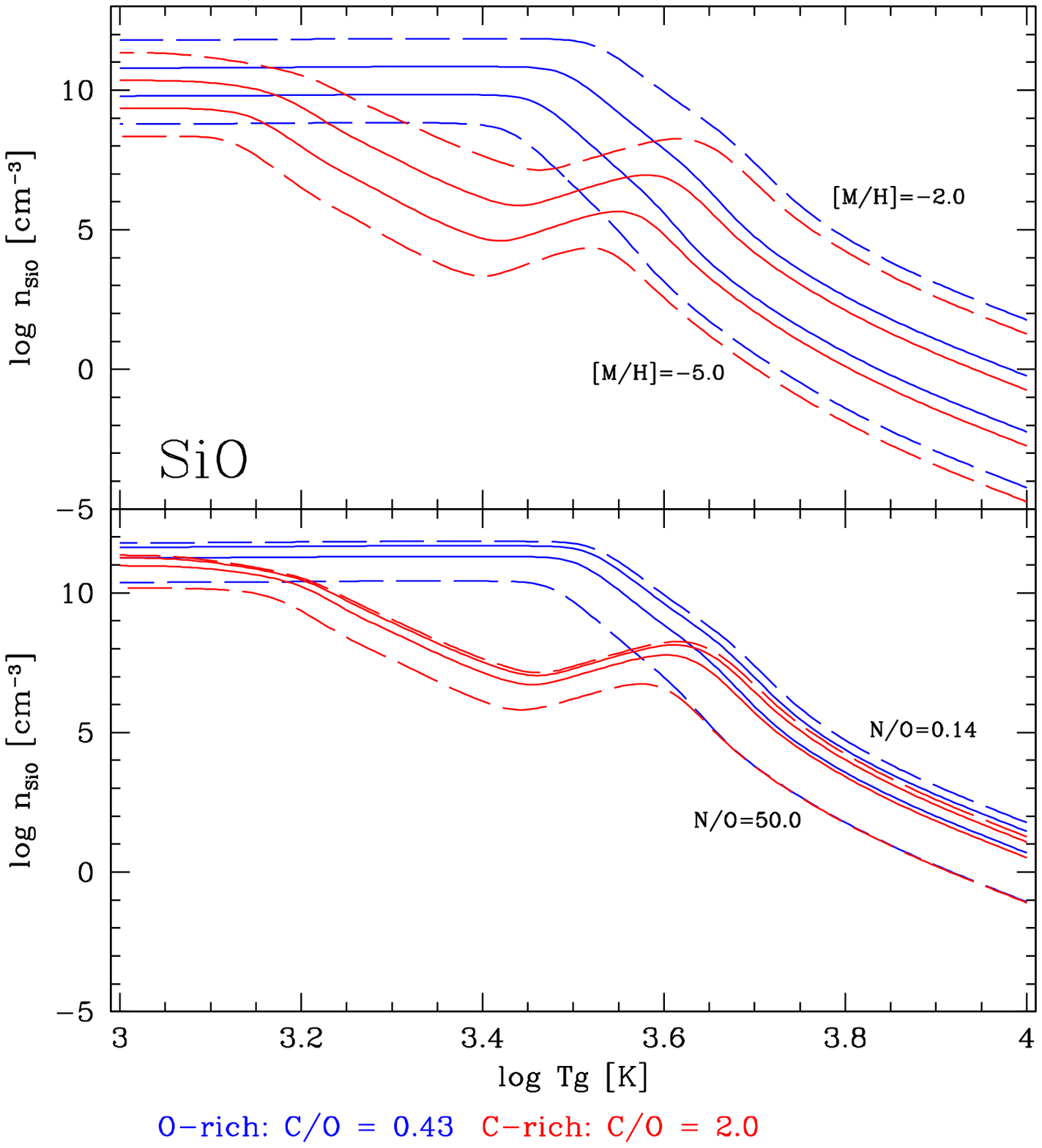}\hspace{-1.2cm}\includegraphics[width=64mm]{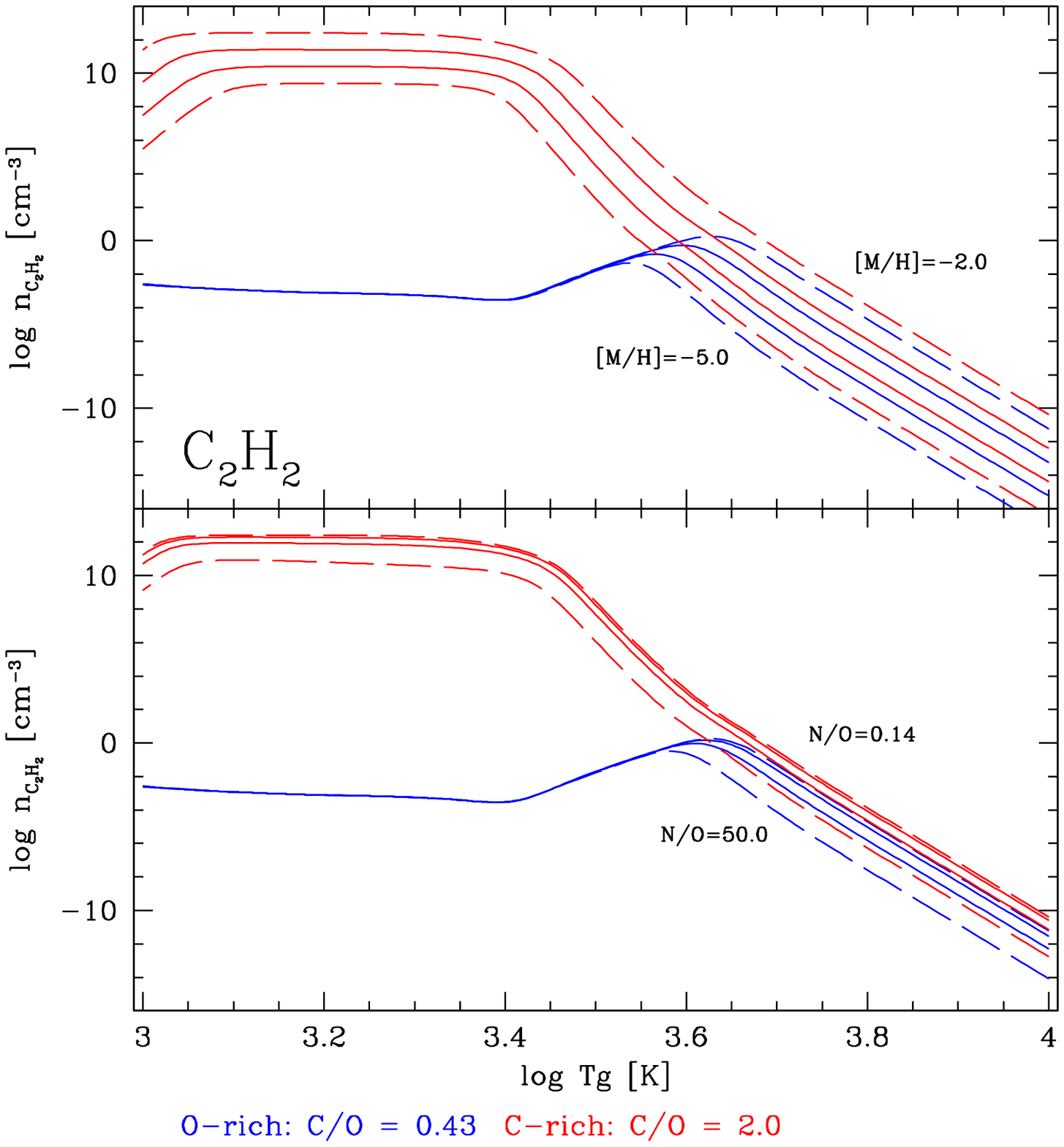}\\
  \hspace{-1cm}\includegraphics[width=64mm]{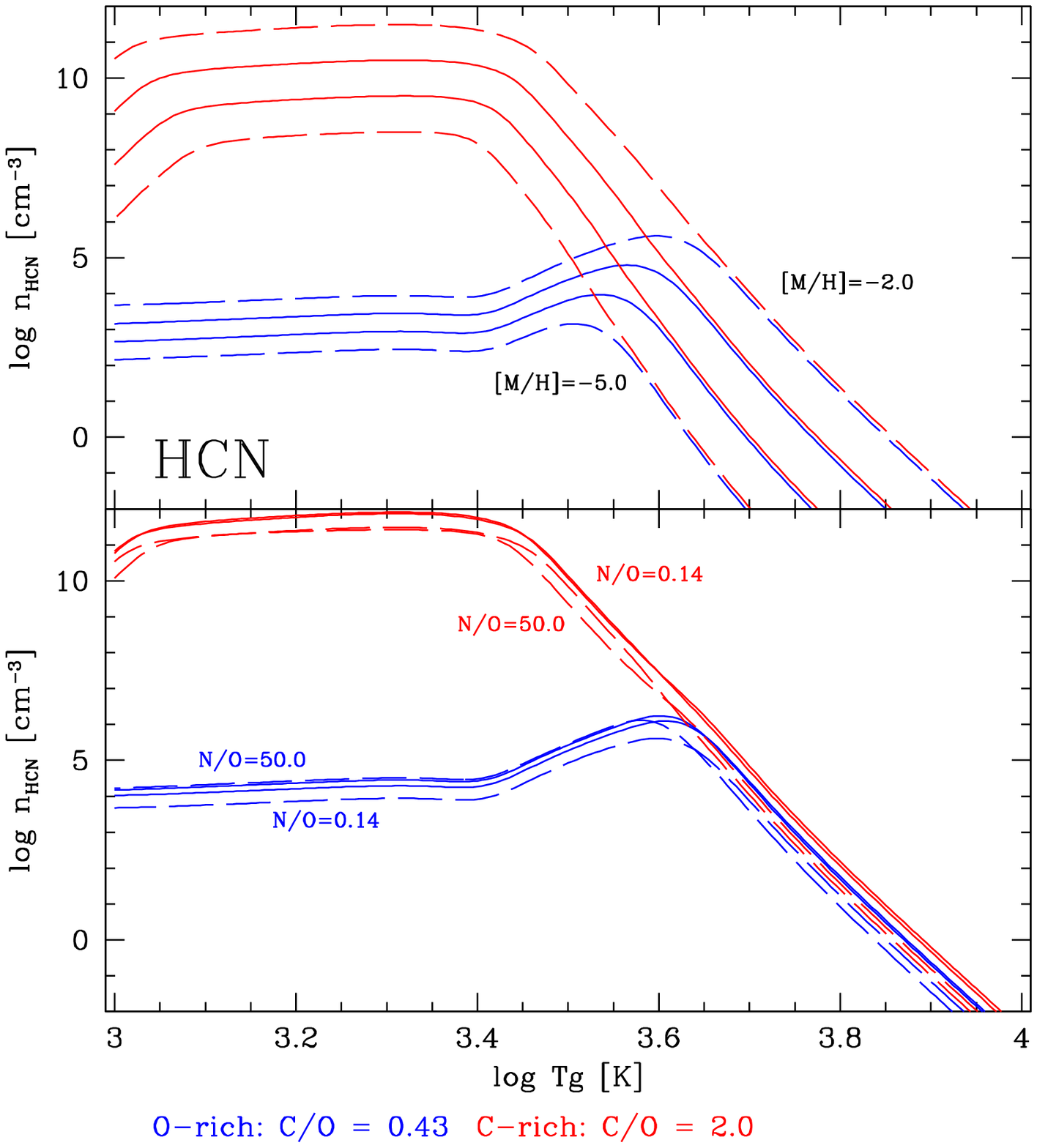}\hspace{-1.2cm}\includegraphics[width=64mm]{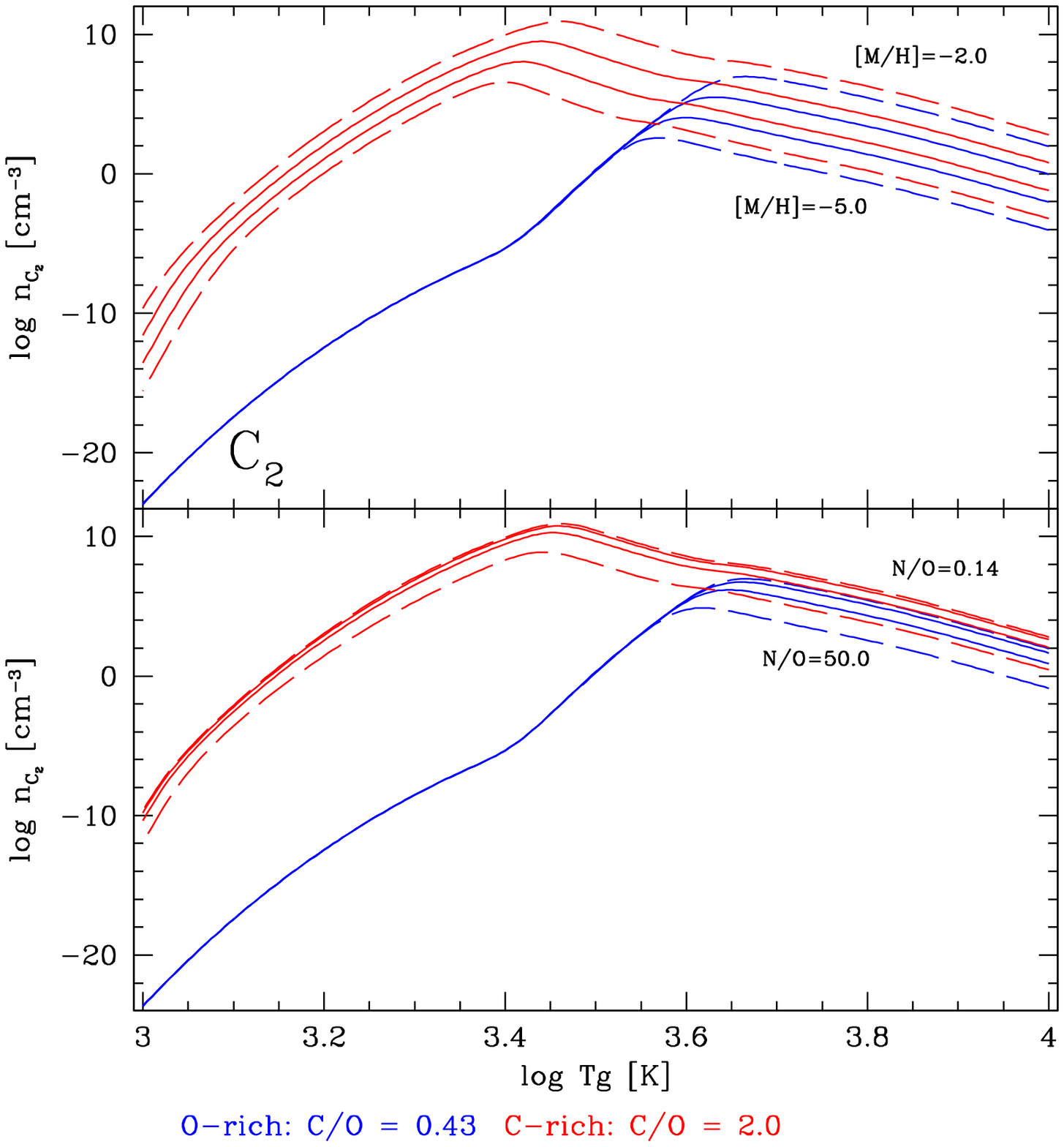}\hspace{-1.2cm}\includegraphics[width=64mm]{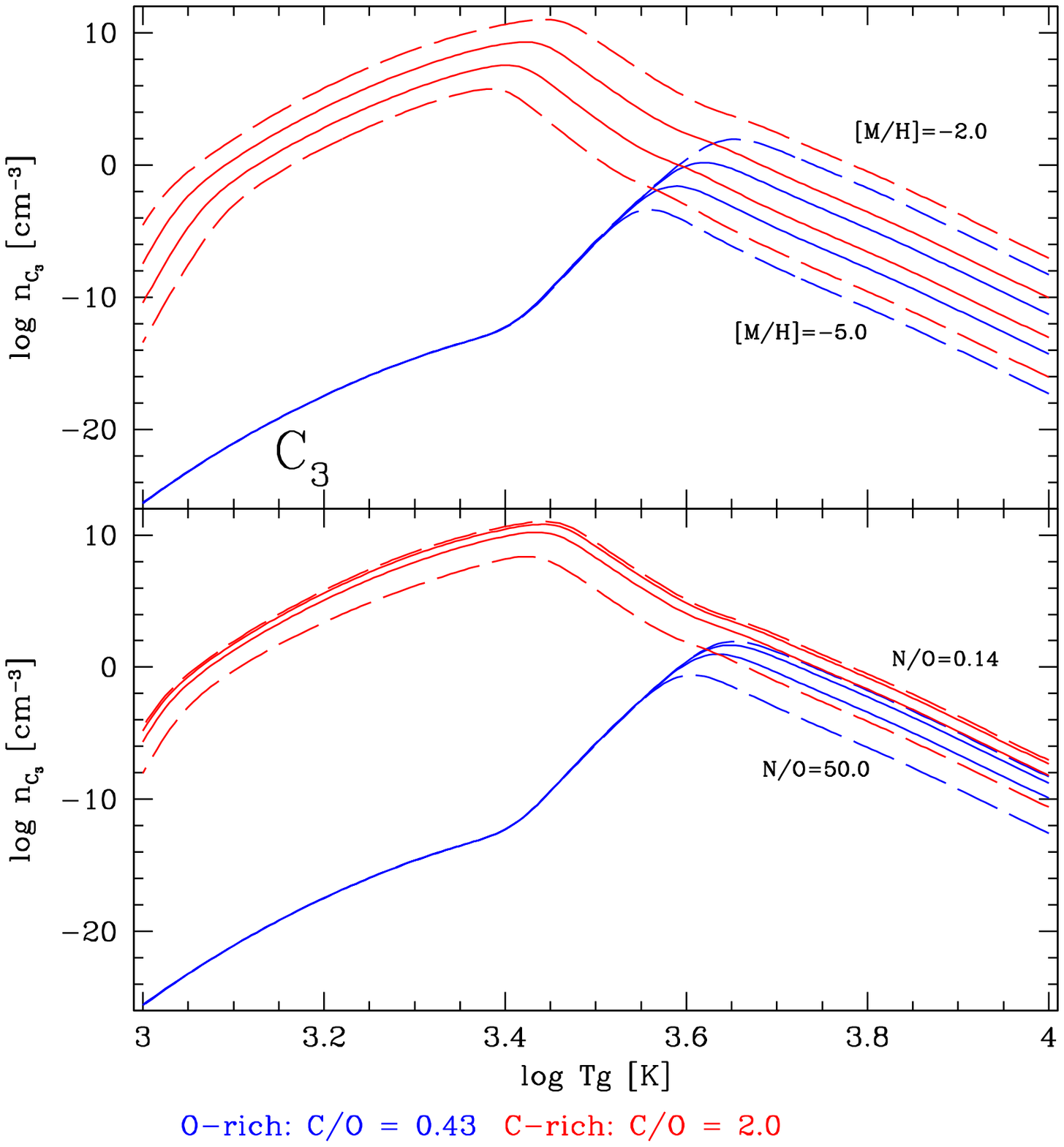}\\
  \caption{Number densities of the line-opacity molecules.  The top
   panels show variation with metallicity, and the bottom variation
   with N/O ratio ($n_\rmn{H} = 1.931 \times 10^{15}$
   $\rmn{cm^{-3}}$).  Dashed lines are labelled with their associated
   values. Solid lines have $[\rmn{M/H}] = 3.0$ and 4.0 and $\rmn{N/O}
   = 1.0$ and 5.0. Each plot contains values for C/O=2.0 (red) and
   C/O=0.43 (blue). Note the scales and ranges of each plot are
   different.}
  \label{chem1}
 \end{figure*}
 
 \begin{figure*}
  \hspace{-1cm}\includegraphics[width=64mm]{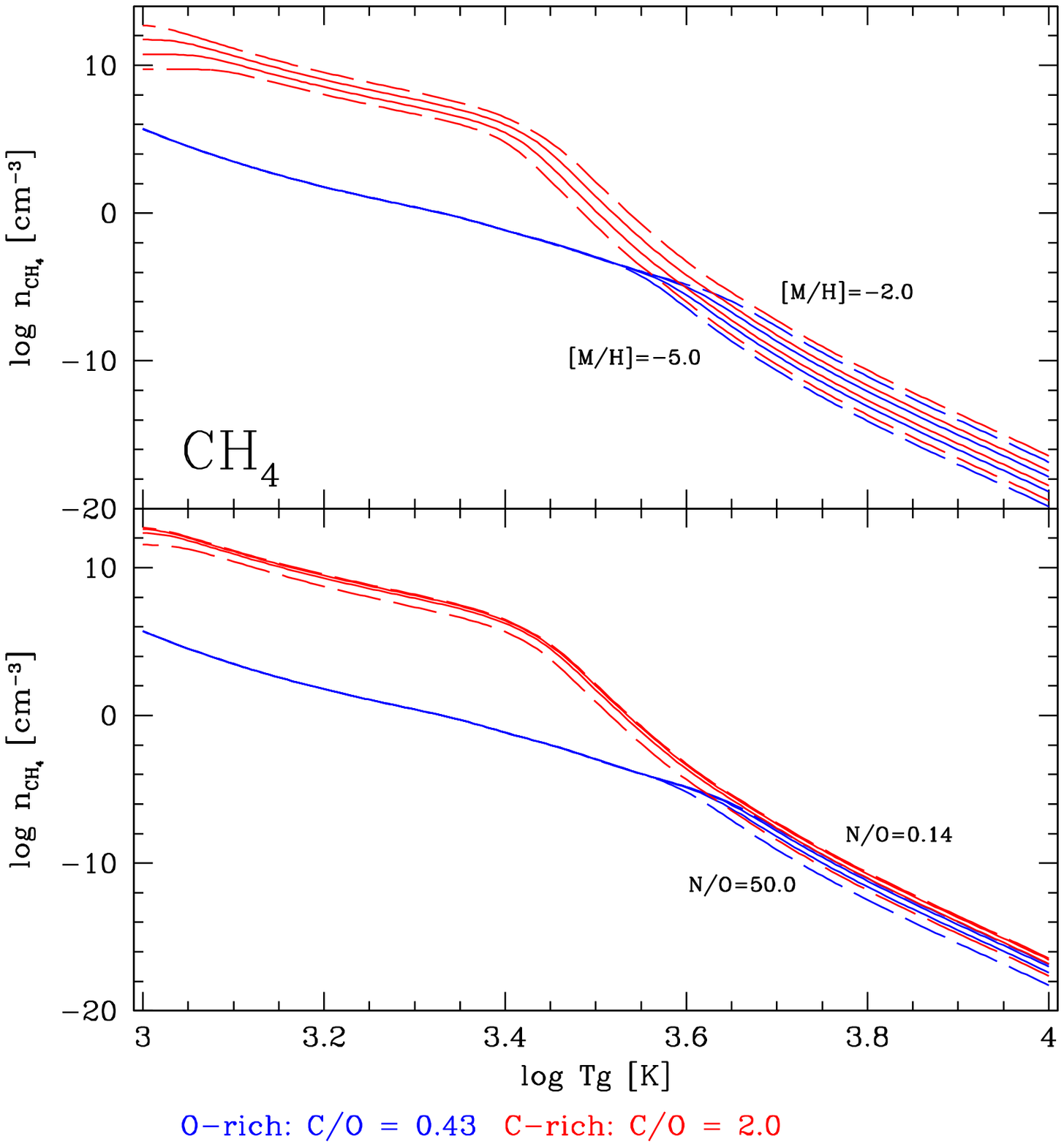}\hspace{-1.2cm}\includegraphics[width=64mm]{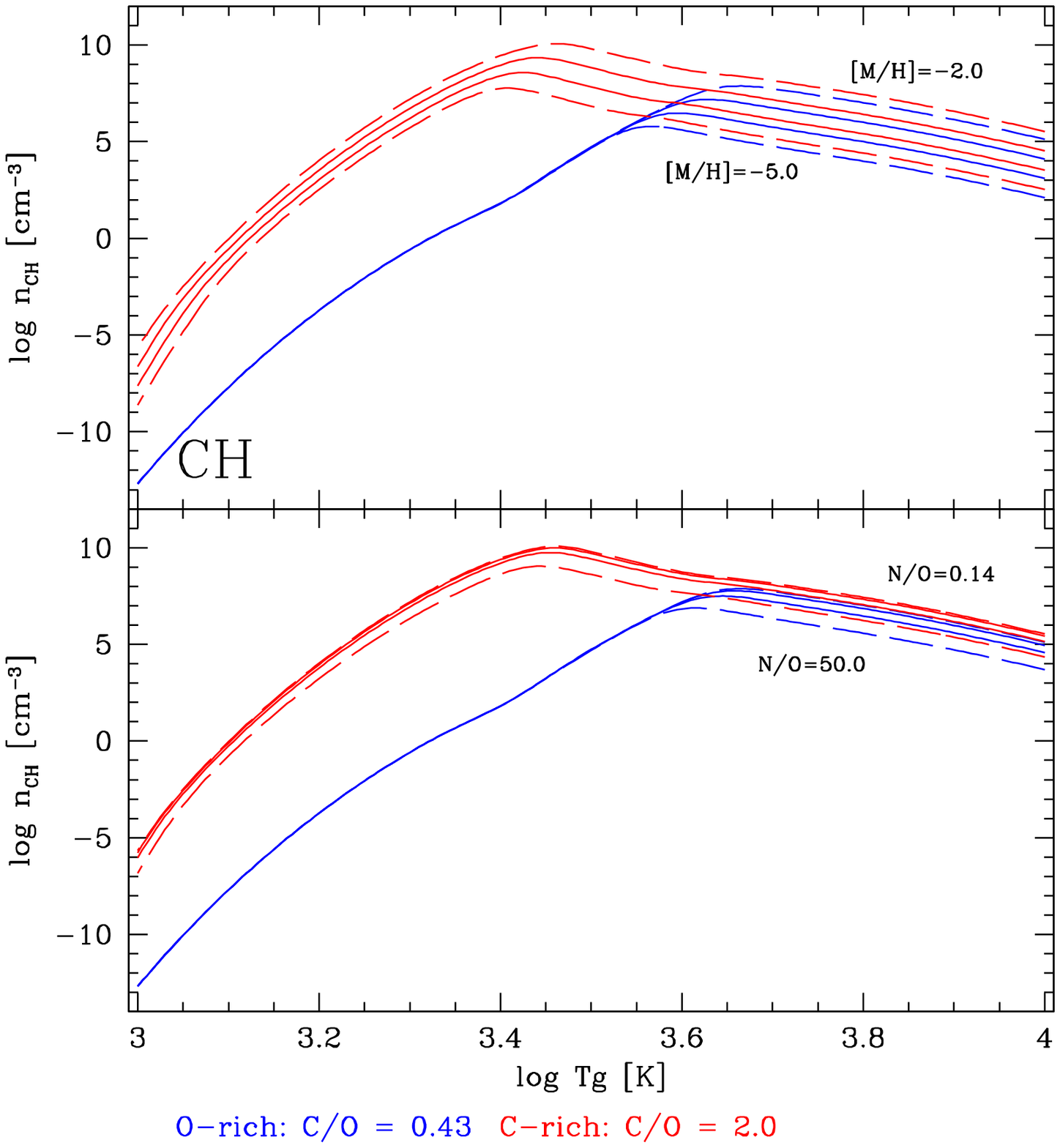}\hspace{-1.2cm}\includegraphics[width=64mm]{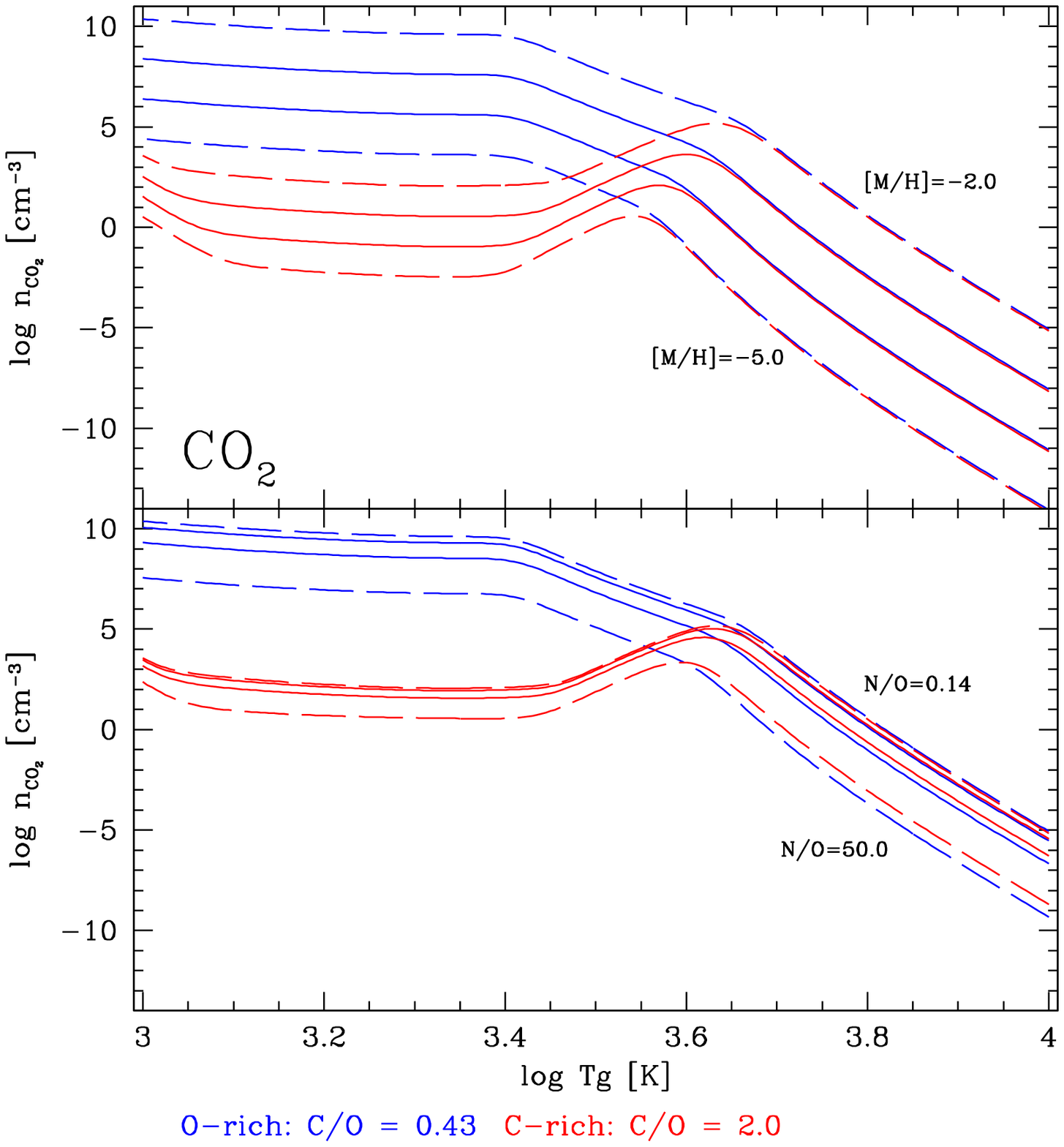}\\
  \hspace{-1cm}\includegraphics[width=64mm]{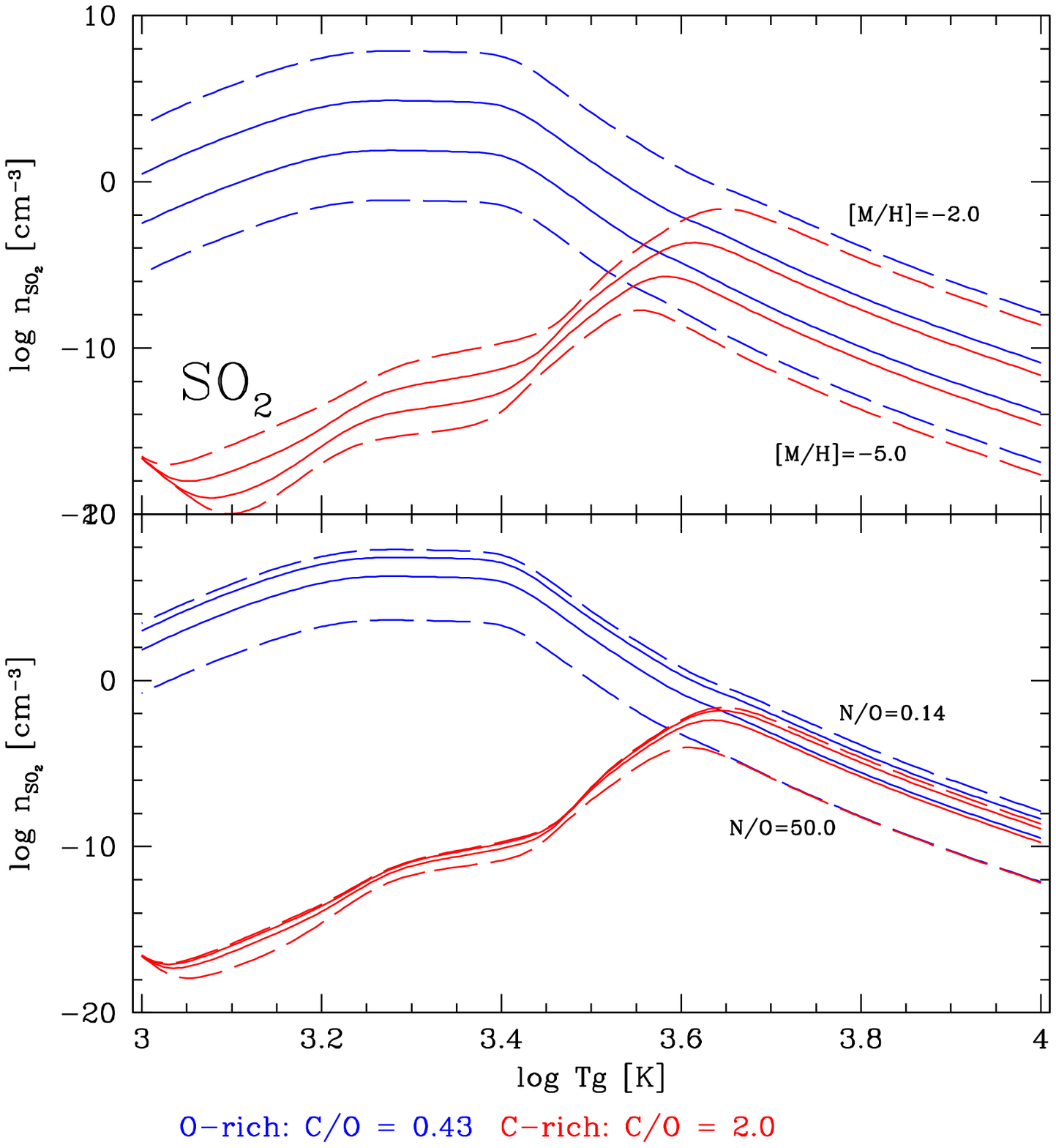}\hspace{-1.2cm}\includegraphics[width=64mm]{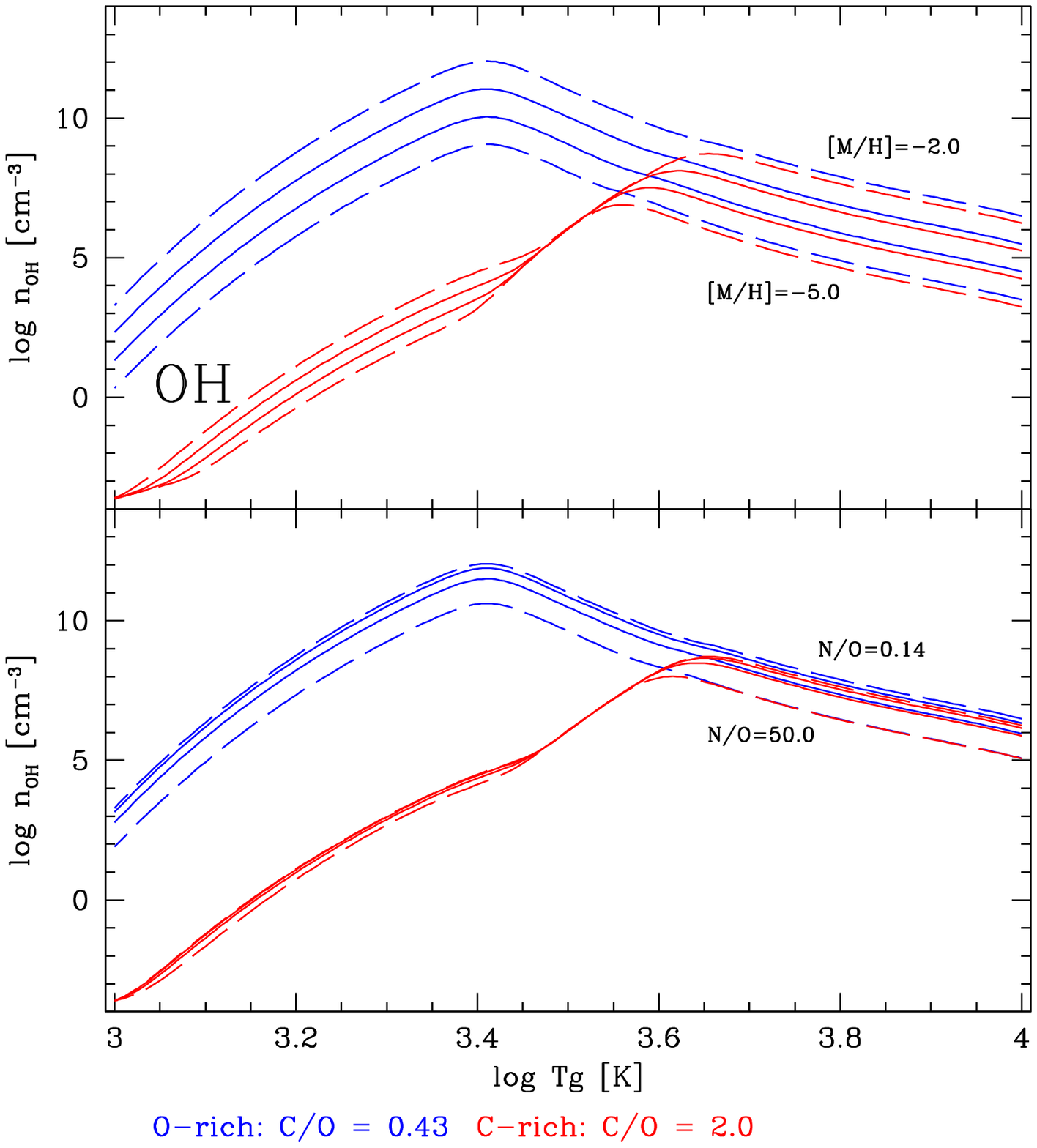}\hspace{-1.2cm}\includegraphics[width=64mm]{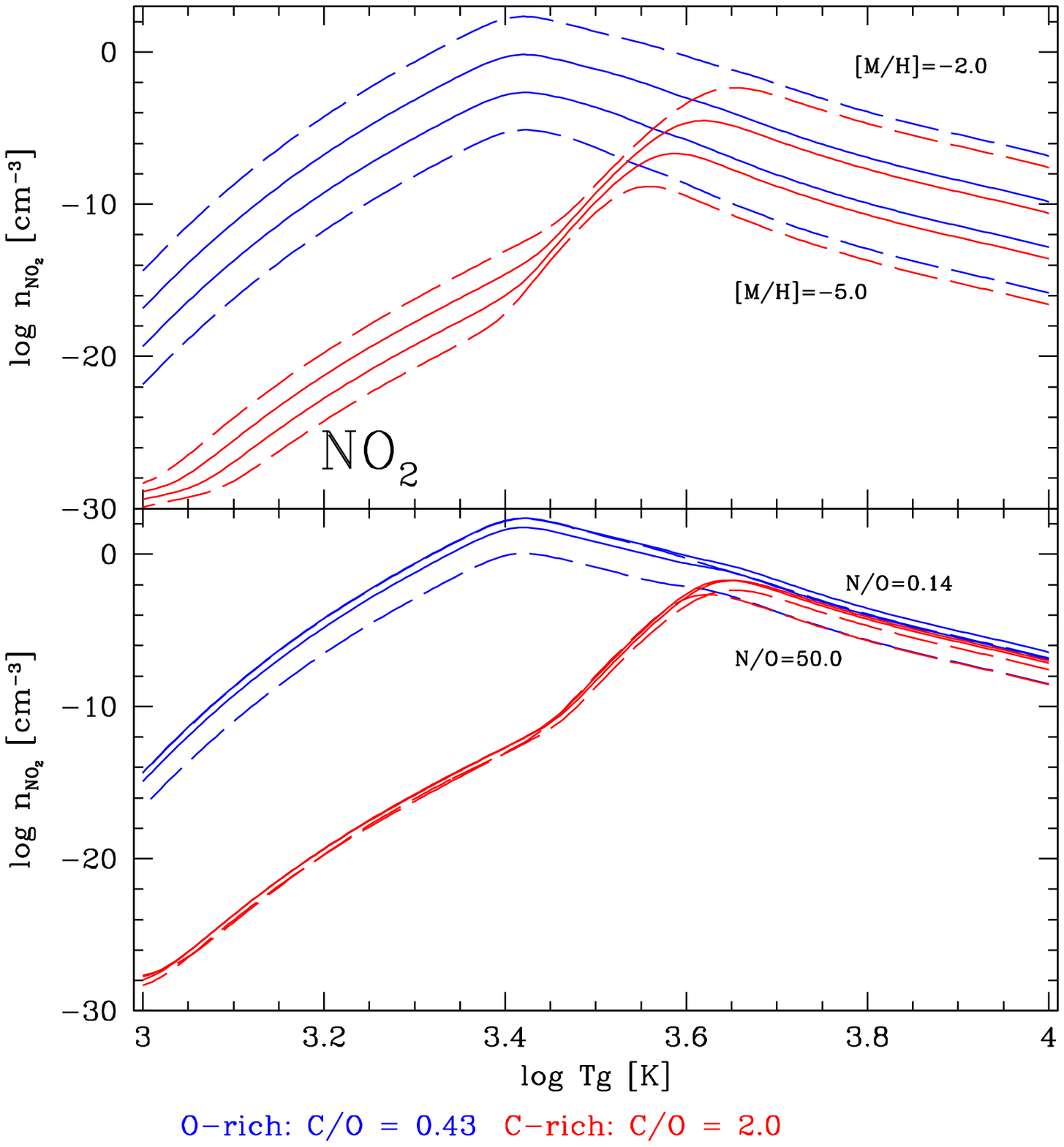}\\
  \hspace{-1cm}\includegraphics[width=64mm]{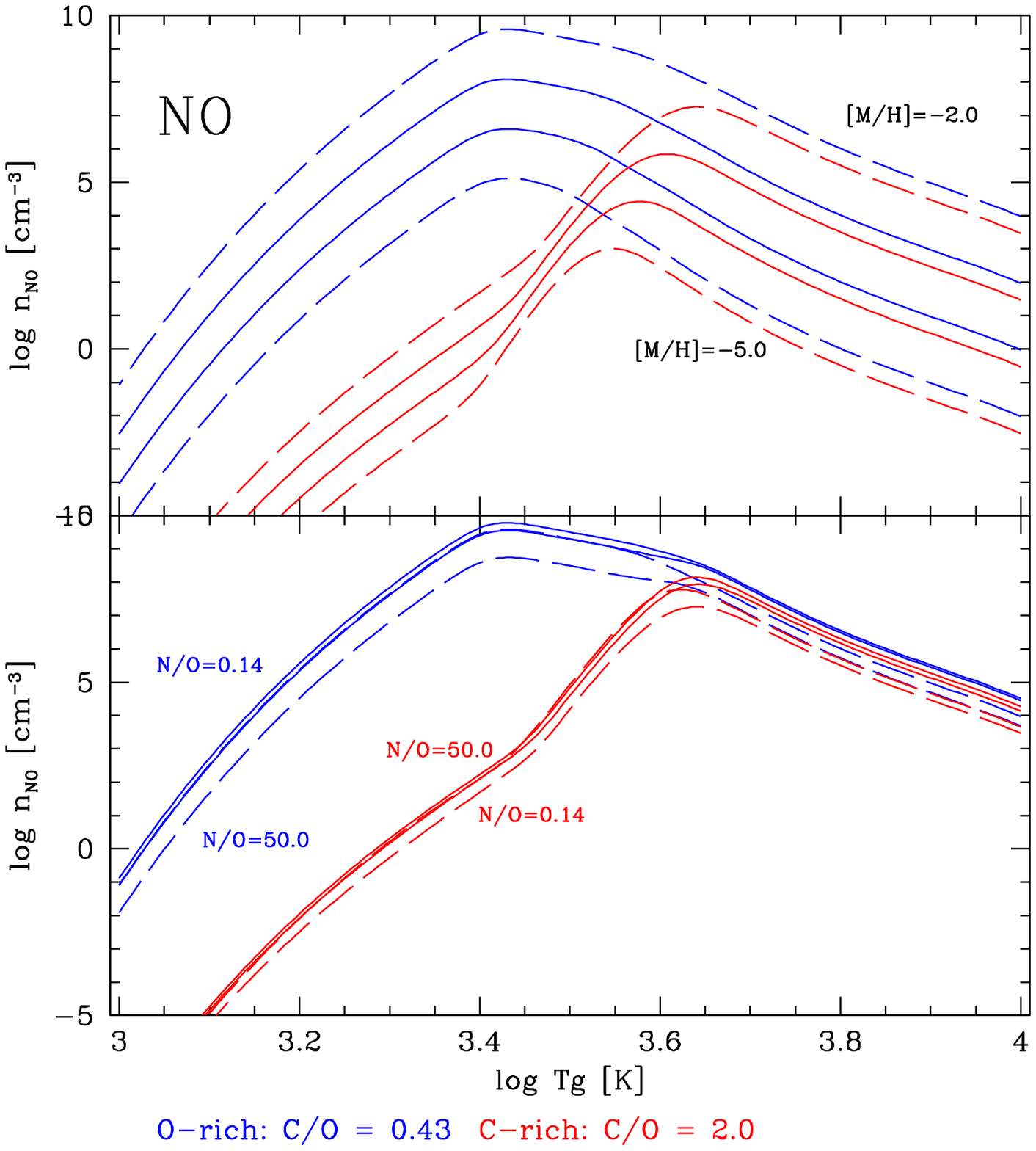}\hspace{-1.2cm}\includegraphics[width=64mm]{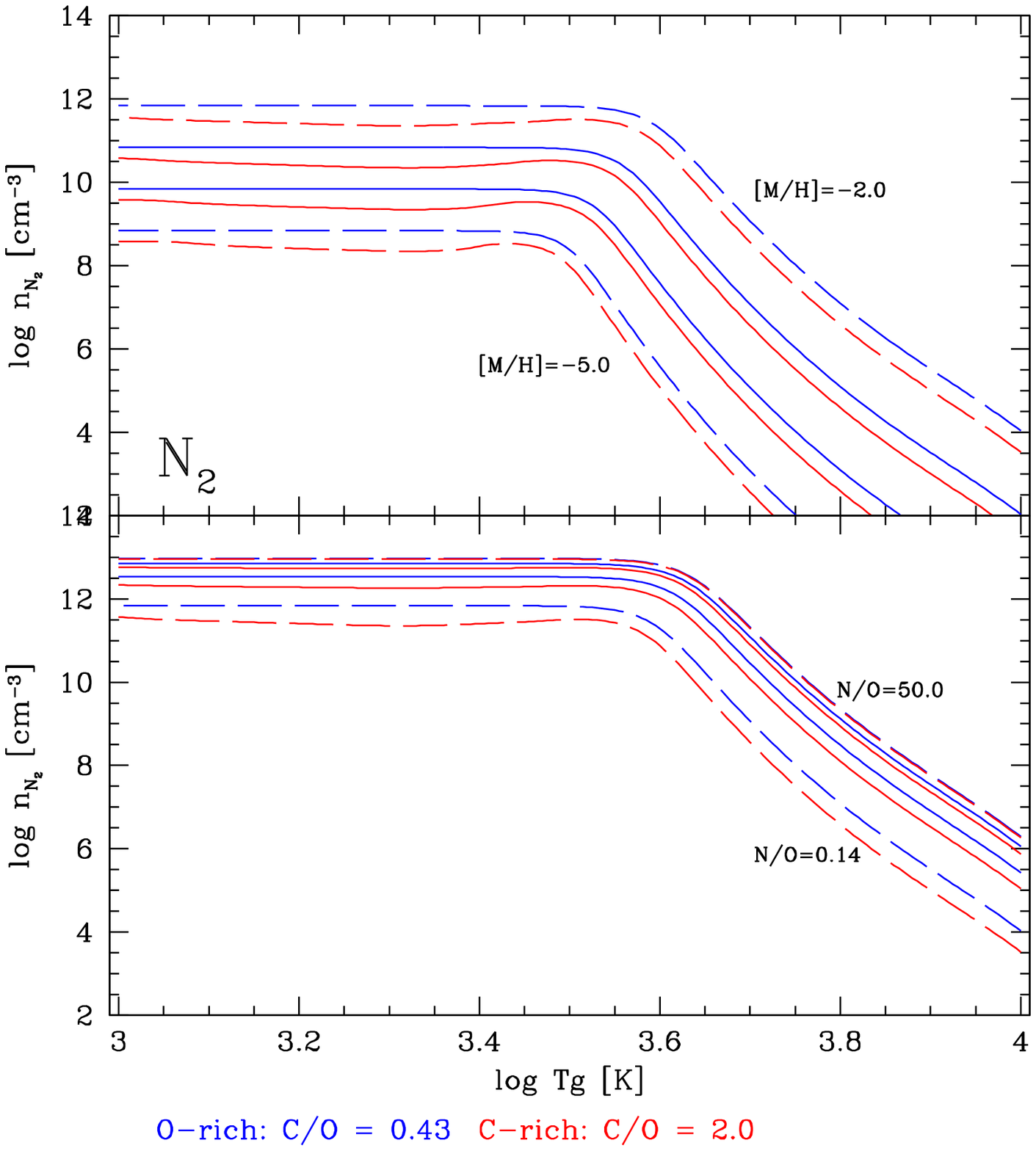}\hspace{-1.2cm}\includegraphics[width=64mm]{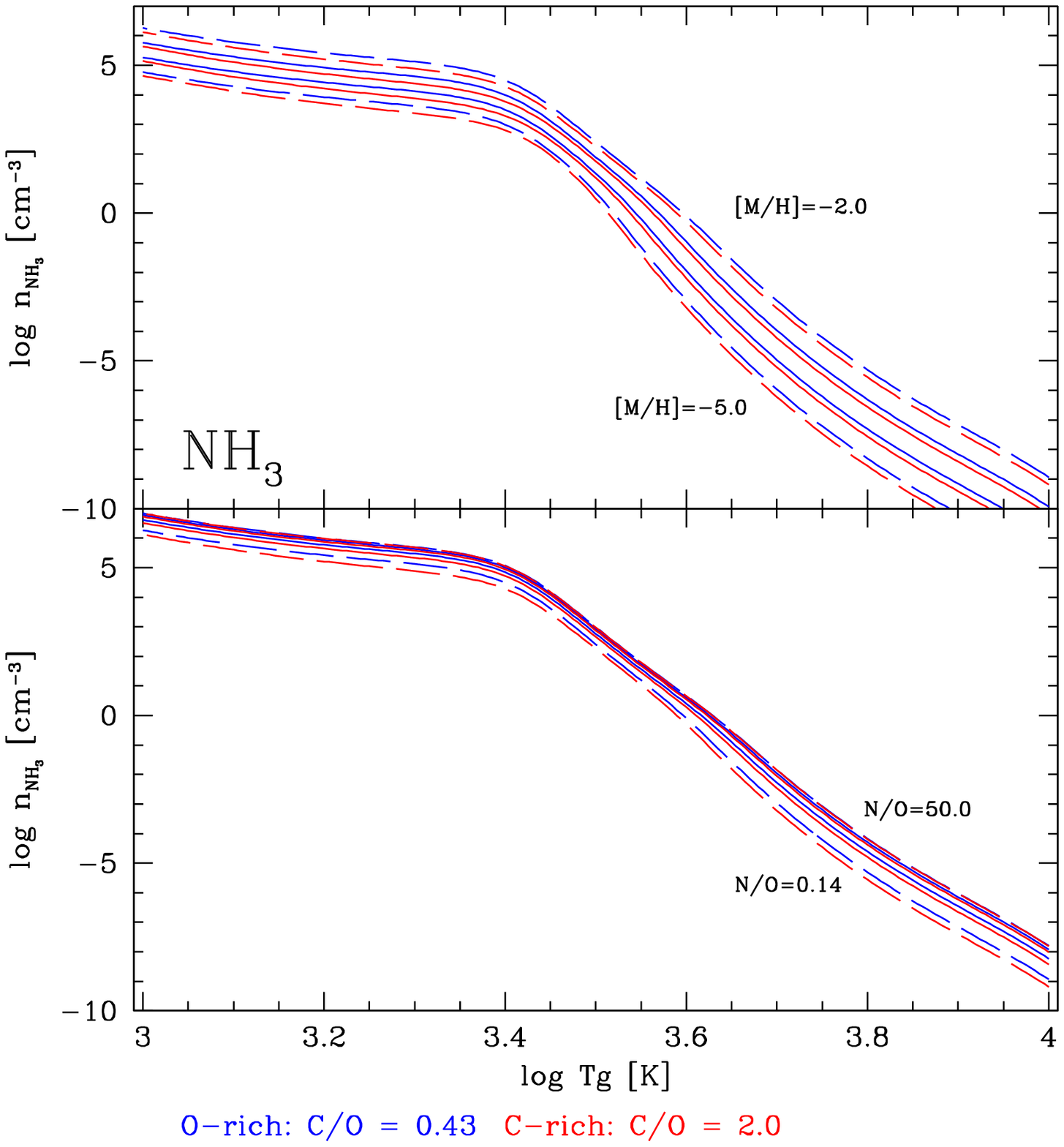}\\
  \caption{Figure~\ref{chem1} is continues here. Several molecules containing nitrogen are
   shown, which demonstrate a dependence on N/O different from the other molecules.}
  \label{chem2}
 \end{figure*}
 
 \begin{figure*}
  \hspace{-1cm}\includegraphics[width=64mm]{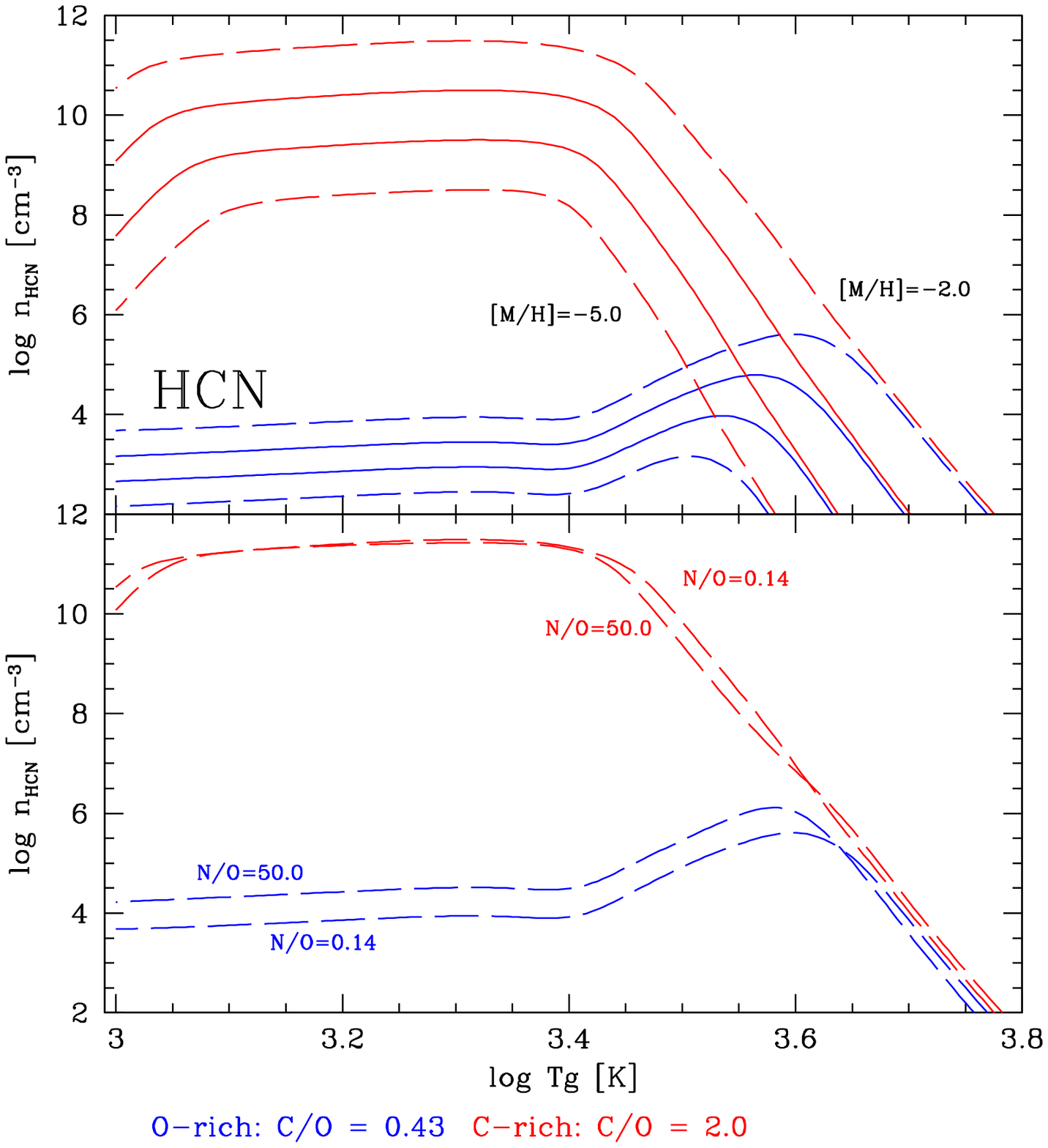}\hspace{-1.2cm}\includegraphics[width=64mm]{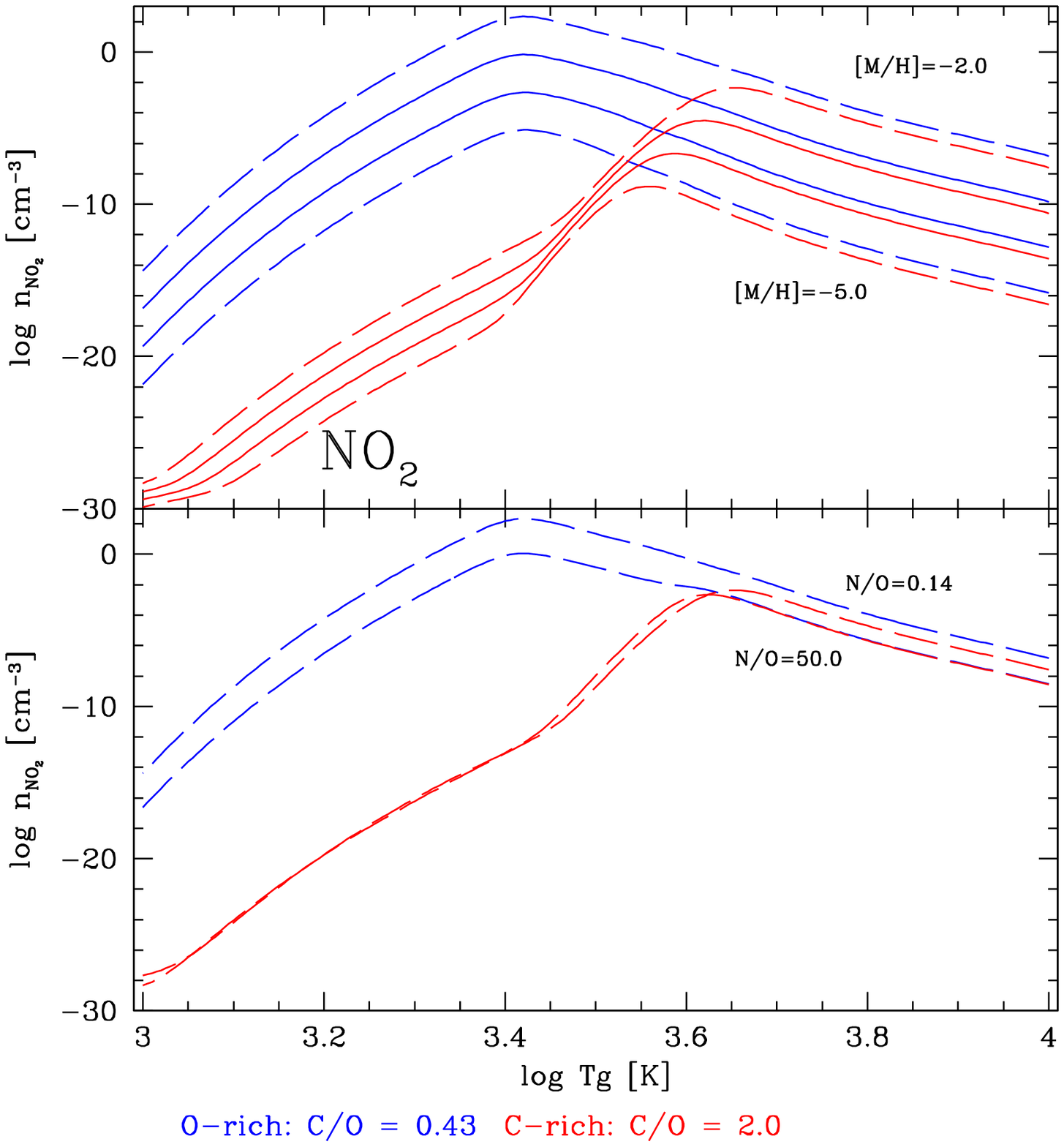}\hspace{-1.2cm}\includegraphics[width=64mm]{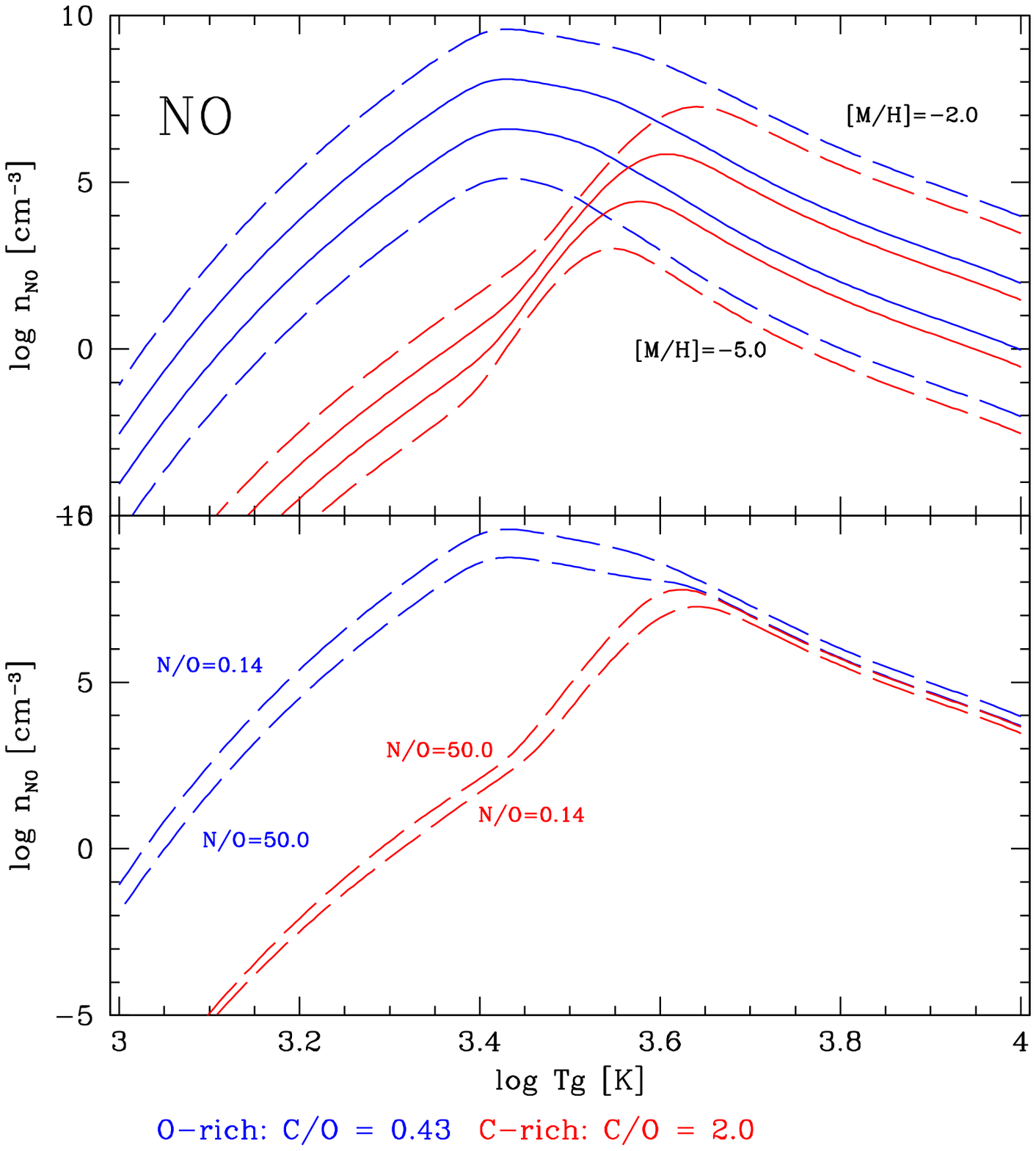}
  \caption{Numbers densities for opacity molecules displaying unusual dependence on N/O. Results are only shown
   for N/O = 0.14 and 5.0, making it easier to see how they vary. HCN and $\rmn{NO}_2$
   show that the lines for each N/O may cross one another several times.}
  \label{chnitro}
 \end{figure*}

\subsubsection{The oxygen abundance}\label{ss:oxygen}

We show how modifying the oxygen abundance affects the resulting
  gas-phase mean opacities. Two sets of opacity tables were made where
  only the oxygen abundance differ.  We test how our mean opacities
  change if we use $\epsilon^{\rm Caffau}_\rmn{O} = 8.76$ suggested by Caffau et
  al. (2008) in comparison to $\epsilon^{\rm GrAsSa}_\rmn{O} = 8.66$ suggested by
  Grevesse, Asplund \& Sauval (2007).  Both values are lower than the
  previously used oxygen abundance in Helling, Winters \& Sedlmayr
  (2000; see their Table A1).  We use the Grevesse, Asplund \& Sauval
  (2007) solar photosphere abundances as specified in
  Table~\ref{abundt} for all other element abundances.  Important to
  note is that while using the Grevesse, Asplund \& Sauval (2007)
  value of $\epsilon^{\rm GrAsSa}_\rmn{O} = 8.66$ the C/O and N/O ratios and the
  metallicity are: $\rmn{C/O} = 0.537$, $\rmn{N/O} = 0.132$,
  $[\rmn{M/H}]_{\odot}^{\rm GrAsSa} = -3.02$. While using the Caffau et al. (2008) values
  $\epsilon^{\rm Caffau}_\rmn{O} = 8.76$, they are: $\rmn{C/O} = 0.427$, $\rmn{N/O}
  = 0.105$, $[\rmn{M/H}]_{\odot}^{\rm Caffau} = -2.97$, simply because adding oxygen means
  adding a metal.

 Figure \ref{caffgasratio} plots for two different gas densities the
  ratio of the mean values for both calculations with different oxygen
  abundances. As expected, the mean opacities for
  $\epsilon_\rmn{O}^{\rm Caffau}$ are higher then for
  $\epsilon_\rmn{O}^{\rm GrAsSa}$ since more oxygen is available to
  form molecules which are opacity sources.  Figure \ref{caffgasratio}
  (top panel) shows that there is a large disparity when changing
  between the different oxygen abundances. While the high-temperature
  Rosseland and Planck mean values are the same since determined by
  atomic/ionic continuum opacities, the deviations can be as large as
  20\%--30\% in the low-temperature regime where molecular line
  absorption dominates the total opacity.  As research into these
  abundances values is ongoing, tables using both values were
  calculated.

\subsubsection{Comparing to earlier calculations}

We compare our new mean opacity tables to tables calculated with
oxygen abundances suggested by earlier works (Fig.~\ref{caffgasratio},
lower panel):\\[-0.5cm]
\begin{tabbing}
$\epsilon^{\rm Lodders}_\rmn{O} = 8.69$  \hspace*{1cm} \= Lodders (2003)\\ 
$\epsilon^{\rm GrNo}_\rmn{O} = 8.87$ \> Grevesse \& Noel (1993)\\
$\epsilon^{\rm Alexander}_\rmn{O} = 8.8093$ \>  Alexander (1975, King
mixture IVa),\\*[-0.7cm]
\end{tabbing}
and to tables from Helling, Winters \& Sedlmayr (2000) for a
collection of Anders \& Grevesse (1989) and Grevesse \& Noel (1993)
abundance data  (Fig.~\ref{old_and_linelist}). The oxygen abundance in these old tables is
$\epsilon^{\rm GrNo}_\rmn{O} = 8.87$ (Grevesse \& Noel 1993). For an
update of Lodders (2003) see Lodders, Palme \& Gail (2008). The lower
panel in Fig.~\ref{caffgasratio} demonstrates that for a given gas
density the $\epsilon^{\rm GrNo}_\rmn{O}$ mean opacities deviate by up
to 40\% in the low-temperature ranges from mean opacities calculate
using the Grevesse, Asplund \& Sauval (2007) atmospheric oxygen
abundance value. Our mean opacity results for the Alexander (1975)
$\epsilon_\rmn{O}$-value fall in-between the results for
$\epsilon^{\rm Lodders}_\rmn{O}$ and $\epsilon^{\rm GrNo}_\rmn{O}$
which is consistent with the comparison shown in Semenov et al.~(2003,
their Fig. 3 left).

The use of the up-dated element abundances by Grevesse, Asplund \&
Sauval (2007) and Caffau et al. (2008), and the use of additional
opacity sources from the HITRAN data base (Sect.~\ref{ss:input}),
causes differences of the present mean opacity values compared to our
earlier results (Helling, Winters \& Sedlmayr 2000). The additional
HITRAN-data mainly affect the mean opacity values below 1000K. 
Figure~\ref{old_and_linelist} (solid blue vs. red dashed) demonstrates
that the maximum total difference is 1 -- 0.5 orders of magnitude at
such low temperatures. Semenov et al. (2003, see their Fig. 2)
present a comparison of our Rosseland and Planck mean gas-opacity data
to other works including Alexander (1975). The influence of the
element abundances on the mean opacities is discussed above with
emphasis on the uncertainty in the oxygen abundances. Note, however,
that all element abundances taken from Grevesse, Asplund \& Sauvale
(2008) are lower then in Helling, Winters \& Sedlmayr (2000, see their
Table A1) including the carbon abundances.

Figure~\ref{old_and_linelist} further suggests that possible uncertainties in
weak-line opacity data can have a considerably larger influence on the
mean opacities than the uncertainties in the element
abundances. However, the purpose of our simple weak-line experiment in
Fig.~\ref{old_and_linelist} was to demonstrate the possible influence on
the mean opacity calculation rather then setting a stringent limit on this
effects.

\subsection{Chemistry}\label{ss:chem}
In order to evaluate the opacity carriers individual abundance
 contribution to the mean opacities, a chemical equilibrium routine
 was run to produce a separate output which gave number densities of
 the various species across the temperature range. The chemistry
 routine considers 158 molecules made of 14 elements, but we only
 present the output for the 18 opacity species included in our mean
 opacity calculation.  Varying the C/O ratio, the N/O ratio, and the
 metallicity, allowed the opacity table's features to be probed
 (Figs. \ref{chem1} - \ref{chnitro}).
 
 The first plot of Fig.~\ref{chem1} shows CO. The density of this
 molecule is effectively constant until $\rmn{log}(T_\rmn{g}) \sim
 3.6$, after which it falls away. This is true in all cases. Only the
 height of the plateau and the exact point at which it falls off vary
 with N/O and metallicity. This shows why the opacity changes so
 strongly between oxygen- and carbon-rich: Of carbon and oxygen, the
 least abundant has many of its atoms locked within CO, making it
 harder to form other molecules containing that element due to its
 high binding energy.
% of 11.1eV.
%Meanwhile, the more abundant element is free to form
% more molecule species.
 
 The result of this is seen for many of the molecules shown in Figs.
 \ref{chem1} and \ref{chem2}, which show large variations between the
 oxygen- and carbon-rich cases. Having $\rmn{C/O} > 1$ result in
 higher densities for the carbon-containing species, and the same is
 true for the case when $\rmn{C/O} < 1$ which affects mostly the
 oxygen-binding molecules.  This is entirely as expected. An
 interesting exception is TiO, where it can be seen that at low
 temperatures ($\rmn{log}(T_\rmn{g}) \sim 3.1$) the densities for the
 carbon-rich case rise above those for the oxygen-rich case, because
 TiO$_2$ becomes more stable than TiO. Lowering the metallicity shifts
 the point where the densities for each case cross; the same occurs
 when increasing N/O. SiO shows similar behaviour, but the rising
 density as temperature decreases flattens just before the high-C
 density overtakes high-O. Both molecules, TiO and SiO, can therefore
 contribute to the absorption of a carbon-rich gas at temperatures below 1000K.
 
 Comparing all the plots, similar features become apparent at certain
 temperatures. In the carbon-rich case, it can be seen that many
 molecules' densities have a turning-point at $\rmn{log}(T_\rmn{g})
 \sim 3.4$. CN, $\rmn{C}_2$ and $\rmn{C}_3$ all reach a peak here
 before falling again, TiO and SiO reach a minimum, and
 $\rmn{C}_2\rmn{H}_2$ and HCN fall off at the end of a plateau.  The
 exact temperature at which these changes occur seems dependent on the
 metallicity: in all cases it appears as though decreasing the
 metallicity shifts the point at which these changes in the molecular
 abundances occur to lower temperatures.
 
 Looking at the opacity tables for $\rmn{C/O} = 2.0$, a peak is
 noticed in both the Planck and Rosseland means at
 $\rmn{log}(T_\rmn{g}) \sim 3.4$. This is caused by the molecules
 whose number densities also reach a maximum here. The peak occurring
 at very low temperatures ($\rmn{log}(T_\rmn{g}) \sim 3.1$) in the
 Rosseland mean could be a result of the changing densities of several
 molecules. At the lower temperature boundary, where the opacity peak
 can be seen to be rising, the TiO and SiO are at their highest points
 in the carbon-rich domain, but immediately fall. At the same time,
 $\rmn{C}_2\rmn{H}_2$ and HCN are increasing to the point where they
 level out into a plateau. Where they reach this point corresponds to
 the peak in the opacity table. While these two molecules stay at
 roughly the same density, the two oxides are removed, lowering the
 opacity.  This holds with Lederer \& Aringer (2008) who found that
 $\rmn{C}_2\rmn{H}_2$ and HCN are the main contributors to their
 carbon-rich Rosseland mean opacities at these low temperatures.
 
 The oxygen-rich molecule densities show a similar `characteristic'
 temperature. At $\rmn{log}(T_\rmn{g}) \sim 3.4$, $\rmn{H}_2\rmn{O}$,
 which is before this point constant, falls off very quickly, while
 TiO reaches a peak. SiO also peaks, but at a point perhaps nearer to
 $\rmn{log}(T_\rmn{g}) \sim 3.5$. As in the carbon-rich case, the
 metallicity shifts the position of the features. The initial plateau
 in the Planck mean and the prominent peak in the Rosseland mean can
 be easily seen to result from the initial high densities of
 $\rmn{H}_2\rmn{O}$, SiO and $\rmn{SO}_2$, and other molecules which
 reach maxima here such as TiO, OH and NO.
 
 In the Planck and Rosseland means for both C/O cases, a small rise
 can be noticed at $\rmn{log}(T_\rmn{g}) \sim 3.6$. This seems to be
 due to an interesting rise in the density of molecules containing the
 least abundant of the two elements, eg. if $\rmn{C/O} > 1$ then
 molecules containing oxygen reach a peak at about this temperature,
 and vice verse. It is important to note that this is the temperature
 beyond which CO's density is seen to fall away, ie.  it begins to
 dissociate. As this happens, atoms of both carbon and oxygen are
 released, allowing other molecules to form. However the rising
 temperature inevitably brings about continued dissociation of these
 molecules as well, leading to the downwards slope seen on all
 plots. Some of the peaks at $\rmn{log}(T_\rmn{g}) \sim 3.6$ are of
 considerable height, allowing contribution the rise noted in the
 opacities.
 
\subsubsection{Changing nitrogen abundance } 

It is also important to note what effects changing the N/O ratio has
 on the molecular densities, and so on the opacities. It can be seen that for
 molecules not containing nitrogen, increasing N/O leads to a
 reduction in the gas density. This is because in order to maintain a
 defined metallicity, when the nitrogen abundance is increased, the
 other metals' abundances must be decreased. For those molecules which
 do contain nitrogen, this is different, and furthermore specific to
 the species in question. For example, the density of NO (Figure
 \ref{chem2}) increases with N/O in the carbon-rich case, while in the
 oxygen-rich case the opposite is true. Figure \ref{chnitro} shows the
 difference more clearly. Meanwhile, $\rmn{NO}_2$ decreases with N/O
 in the oxygen-rich case, yet when $\rmn{C/O} = 2.0$ the relationship
 becomes very unclear with a seeming dependence on temperature as well
 as N/O. Despite these more complex relationships, most molecules
 still decrease in number with increasing N/O, leading to the overall
 decrease in the mean opacities observed in Figure \ref{tables}.
 
 Comparing the plots of HCN and $\rmn{NO}_2$ in Figures \ref{chem1}
 and \ref{chem2} with their counterparts in Figure \ref{chnitro}, it
 may be seen that in the carbon-rich cases, and at
 $\rmn{log}(T_\rmn{g}) \ga 3.5$ when oxygen-rich, the molecules'
 number densities are lower at N/O = 0.14 and 50.0 than they are when
 N/O = 1.0 or 5.0. This suggests an optimal N/O ratio which is
 associated with a maximum in the number density of
 nitrogen-containing molecules, and dependent on competition between
 the abundances of carbon, nitrogen and oxygen. Also note that
 $\rmn{NO}_2$ remains in very low numbers throughout the whole
 temperature range. While this molecule has little effect on the mean
 opacities in the temperature interval considered here
 ($10^3\,\ldots\,10^4$K), HCN in particular has a strong influence.
 
 Finally, also included in the opacity calculations were lines by
 collision induced absorption (CIA) by $\rmn{H}_{2}\rmn{-H}_{2}$ and
 $\rmn{H}_2\rmn{-He}$. These are known to contribute significantly to
 lower temperature and high pressure opacities, however due to the
 nature of this kind of absorption it cannot be probed as has been
 done for the molecular lines. Borysow, J\o rgensen \& Zheng (1997)
 demonstrated that collision induced absorption is one of the most
 important opacity source in the atmospheres of the lowest metallicity
 stars (their Fig.~6).

\section{Summary}
 We present a set of gas-phase Planck mean and Rosseland mean opacity
tables applicable for simulations of star and planet formation,
stellar evolution, disk modelling at various metallicities of a
hydrogen-rich gas. The Rosseland mean values are suitable in hot and
dense regions of an absorbing gas, where the mean free path of all
photons of arbitrary wavelength is smaller than the characteristic
scale height of the gas, and hence, the radiative transfer diffusion
approximation holds. Here, the large number of weak absorption lines
is well represented by the Rosseland mean opacity values. If the gas
becomes optically thin, the diffusion approximation becomes invalid,
and single, strong absorbing frequencies may become important and
determine the radiative flux. In the limiting case of a missing
correlation between the frequency distribution of the flux and the
frequency distribution of the opacity, e.g. due to line shifts induced
by velocity fields, the Planck mean opacity values can serves as a
good approximation.

Tables are produced for C/O= 0.43$\,\ldots\,100.0$, for N/O=
0.14$\,\ldots\,100.0$, and [M/H]$^{\prime}=\log N_{\rm M}/N_{\rm
  H}=-2\,\ldots\,-7.0$ ([M/H]=[M/H]$^{\prime}-$[M/H]$_{\odot}$,
[M/H]$_{\odot}\approx -3$, see Sect.\ref{ss:oxygen}).  We have shown
that the present uncertainty regarding the oxygen abundance values do
affect the mean opacity results. We therefor provide tables for the
values from Grevesse, Asplund \& Sauval~(2007) and for the value from
Caffau et al.(2008). We demonstrate the dependence of our mean opacity
values on C/O, N/O and metallicity. The oxygen-to-carbon-rich
transition range around C/O$\approx 1$ is particular interesting, and
our results demonstrate that weak line opacity sources are important
here. We caution, however, that uncertainties in the weak-line
  opacity data can strongly influence the Rosseland mean opacities at
  a tentative threshold of $\lesssim 10^{-2}$cm$^2$/g for $\kappa^{\rm
    i}(\lambda)$ in Eq.~\ref{eq:Ross}.

 The opacity tables will be available online under
http://star-www.st-and.ac.uk/$\sim$ch80/datasources.html.

\section{Acknowledgments}
We thank the referees for their careful reading and valuable
suggestion which considerably improved the manuscript. WL thanks the St
Andrews Physics Trust for funding his summer placement at the
University of St Andrews.  ChH thanks Adam Burrows for sharing his
chemical equilibrium constants data for the molecular FeH and CO. The
computer support at the School of Physics and Astronomy in St Andrews
is highly acknowledged.

% References ie. bibliography

\end{document}